\documentclass[trackchanges,twocolumn]{aastex62}
\usepackage{natbib}
\usepackage{amssymb,amsmath}
\usepackage{graphicx}
\usepackage{multirow}
\usepackage{color}
\usepackage{hyperref}
\usepackage{fancyhdr}
\usepackage[T1]{fontenc}
\usepackage{float}
\usepackage{xcolor,colortbl}
\UseRawInputEncoding

\newcommand{\sslucb}{\affiliation{Space Sciences Laboratory, University of California, Berkeley, CA 94720-7450, USA}}

\newcommand{\uofa}{\affiliation{ Lunar and Planetary Laboratory, University of Arizona, Tucson, AZ 85721, USA}}
\newcommand{\uofap}{\affiliation{Department of Physics, University of Arizona, Tucson, AZ 85721, USA}}

\newcommand{\lesia}{\affiliation{LESIA, Observatoire de Paris, Universit\'e PSL, CNRS, Sorbonne Universit\'e, Universit\'e de Paris, 92195 Meudon, France.}}

\newcommand{\revise}[1]{{#1}}

\begin{document}

\title{Estimated Heating Rates Due to Cyclotron Damping of Ion-scale Waves Observed by Parker Solar Probe }

\author[0000-0002-8941-3463]{Niranjana Shankarappa}\uofap
\author[0000-0001-6038-1923]{Kristopher G. Klein}\uofa
\author[0000-0002-7365-0472]{Mihailo M. Martinovi\'c}\uofa \lesia
\author[0000-0002-4625-3332]{Trevor A. Bowen}\sslucb


\begin{abstract}
 Circularly polarized waves consistent with parallel-propagating ion cyclotron waves (ICWs) and fast magnetosonic waves (FMWs) are often observed by Parker Solar Probe (PSP) at ion kinetic scales. Such waves damp energy via the cyclotron resonance, with such damping expected to play a significant role in the enhanced, anisotropic heating of the solar wind observed in the inner heliosphere. We employ a linear plasma dispersion solver, \texttt{PLUME}, to evaluate frequencies of ICWs and FMWs in the plasma rest frame and Doppler-shift them to the spacecraft frame, calculating their damping rates at frequencies where persistently high values of circular polarization are observed. We find such ion-scale waves are observed during $20.37\%$ of PSP Encounters 1 and 2 observations and their plasma frame frequencies are consistent with them being transient ICWs. We estimate significant ICW dissipation onto protons, consistent with previous empirical estimates for the total turbulent damping rates, indicating that ICW dissipation could account for the observed enhancements in the proton temperature and its anisotropy with respect to the mean magnetic field.
\end{abstract}

\section{Introduction}
Understanding the mechanisms that drive solar coronal heating and the acceleration of the solar wind is a long-standing goal of space plasma physics. This goal requires the identification of the processes that couple electromagnetic fields to the protons and electrons. Dissipation of waves and turbulent fluctuations \citep{Matthaeus_1999} can occur via many mechanisms that can broadly be grouped into three classes: resonant mechanisms (e.g. Landau and transit time damping, \cite{Barnes_1966}, \cite{Leamon_1999}, \cite{Chen_2019_LD}), non-resonant mechanisms (e.g. stochastic heating, \cite{Chen_2001_SH}, \cite{Chandran_2010}), and intermittent dissipation in current sheets and magnetic reconnection sites \citep{Dmitruk_2004, Matthaeus_Velli_2011}. Multiple dissipation mechanisms can operate in the same environment \citep{He_2015, Klein_2020_JPP_Ld_Cyc_FPC} making it necessary to estimate the damping associated with each individually. 

Cyclotron damping is a plausible wave-particle resonant mechanism for damping circularly polarized waves, whose presence in the solar wind is supported by spacecraft observations \citep{Leamon_1998_ICW, Kasper_2013_ICW_damping, Bowen_2022_Cyclotron_dissipation}. 
Cyclotron damping of ion cyclotron waves (ICWs) has been proposed to be a significant process leading to the anisotropic heating of protons that enhances at close radial distances from the Sun \citep{Hollweg_Isenberg_ICW_diss}. 
Ion-scale waves are ubiquitous in the young solar wind \citep[e.g.][]{Bowen_ICW, Verniero_2020, Zhao_2021, Shankarappa_2023} and at 1 AU \citep[e.g.][]{Jian_2009, Woodham_2021}. 

Parker Solar Probe (PSP; \cite{Fox:2015}) has been making in-situ measurements of the solar wind thermal plasma \citep{Kasper2016} and electromagnetic fields \citep{Bale2016} at unprecedented distances from the Sun since its' launch in 2018. 
In our previous work, \cite{Shankarappa_2023} (referred to as SKM23 hereon), we analyzed PSP observations from Encounters 1 and 2 between heliocentric distances 0.166 and 0.25 au and inferred the feasibility of Landau damping to dissipate turbulent fluctuations. We further inferred the presence of significant energy in circularly polarized ion-scale waves--consistent with parallel propagating (with respect to the local magnetic field) ICWs or fast magnetosonic waves (FMWs)--coincident with wavevector-anisotropic background turbulence. 

In this work, we expand upon that previous work by estimating the damping rates of ion-scale wave energy onto protons and electrons via the parallel cyclotron resonance. Such an estimation requires the identification of the observed waves as ICWs or FMWs. \cite{Bowen_Outward_ICW_2020} have found PSP electric field observations on November 4, 2018 to be consistent with outward propagating ICWs and FMWs. However, the accurate processing of the full electric field vector measurement by PSP across entire encounters is limited by the geometry of the antennas. Due to this limitation and inherent difficulties with uniquely determining the plasma frame polarization of waves using just magnetic field observations from a single spacecraft, the observed ion-scale waves cannot be generally uniquely distinguished as ICWs or FMWs. To compensate for this uncertainty, we employ the \texttt{PLUME} linear plasma dispersion solver \citep{Klein_Howes_2015_PLUME} to estimate the damping rates of the observed waves as both ICWs or FMWs.

We find that consistent with \cite{Bowen_ICW} and \cite{Liu_2023}, the ion-scale waves are observed during $20.37\%$ of PSP Encounters 1 \& 2 solar wind observations. The plasma frame frequencies corresponding to the frequencies where the waves are observed are consistent with them being locally generated ICWs. We infer a significant possible ICW dissipation onto protons that could explain the observed enhancement in the proton temperature anisotropy at closer distances to the Sun \citep{Mozer_2023, Zaslavsky_2023}. The heating rates estimated in this work are in very good agreement with the empirical estimates of total turbulent dissipation rates by \cite{Bandyopadhyay_2020}. 

The rest of this paper is organized as follows. The methodology employed to evaluate heating rates via cyclotron damping of ion-scale waves is described in Section~\ref{sec:methodology}. In Section~\ref{sec:occurence} we quantify the duration and energy of ion-scale waves observed in Encounters 1 and 2. We further discuss their transience and compare their plasma frame frequencies to frequencies over which ICWs and FMWs would be generated due to proton temperature anisotropy instabilities. Section~\ref{sec:dissipation} presents the proton heating rates estimated in this work and the ratio of proton to total heating rates estimated by including previous work, followed by a discussion in Section~\ref{sec:discussion}.

\begin{figure*}
    \centering
    \plotone{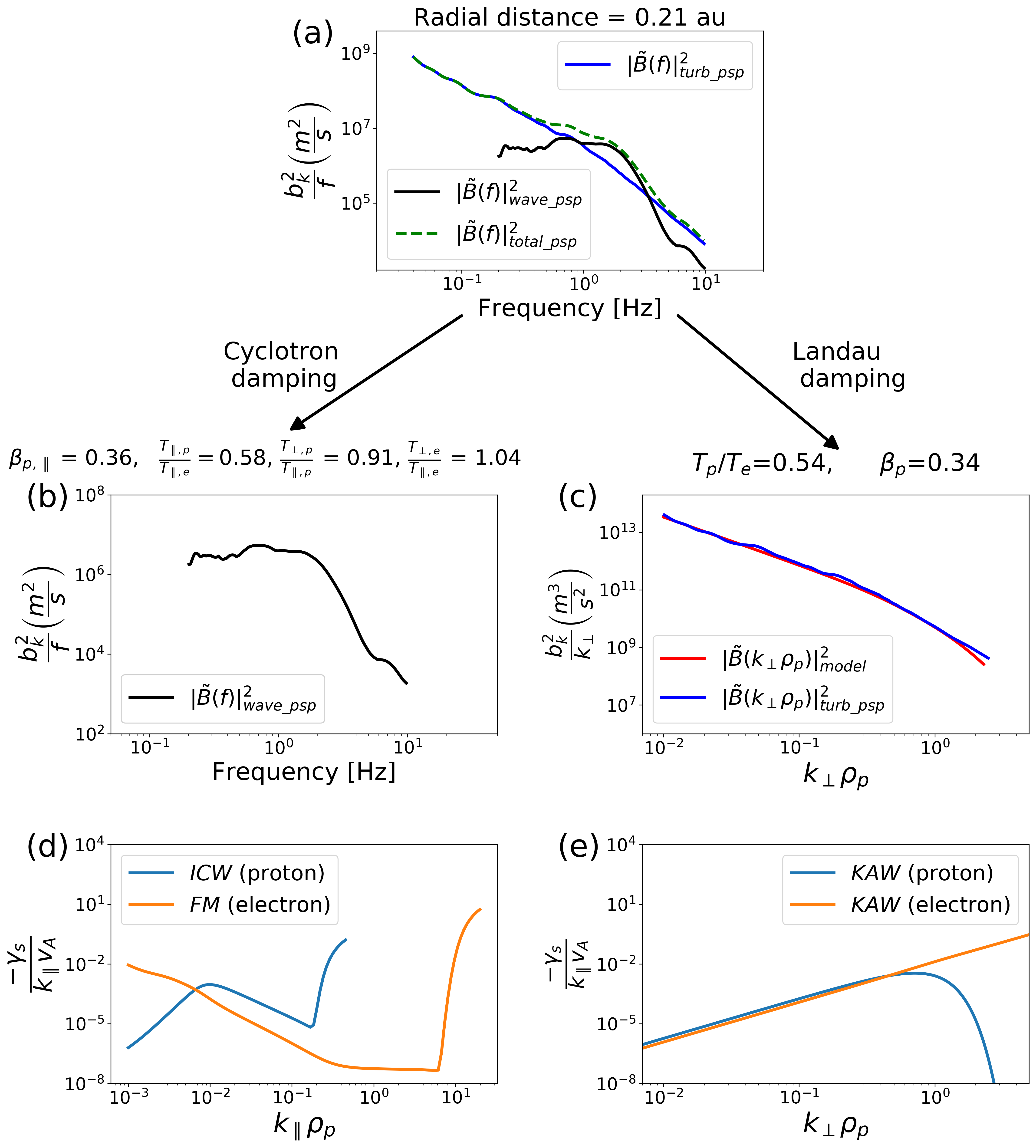}
    \caption{Overview of the estimation of turbulent and ion-scale-wave dissipation for an example 15-minute PSP interval on 10th November 2018, 17:53:58 - 18:08:53,  where ion-scale waves are observed: (a) The ion-scale wave energy is distinguished from turbulent energy spectrum. (c,e) The dissipation rates of turbulent energy via the Landau resonance were estimated using a 1-D cascade model (SKM23). (b,d) The dissipation rates of ion-scale wave energy via the cyclotron resonance as estimated in this work. 
    }
    \label{fig:wave_turb_diss_master_plot}
\end{figure*}
\section{Methodology} \label{sec:methodology}
 
The heating rates due to cyclotron damping of ion-scale waves are evaluated using the following steps for each 15-minute interval of PSP observations analyzed in \cite{Shankarappa_2023} (SKM23).

\subsection{Plasma and Magnetic Field Data}
 We use Level 2 magnetic field data (in spacecraft frame RTN coordinates) from the Flux Gate Magnetometer (MAG) component of the PSP/FIELDS instrument suite, which makes in situ observations of electromagnetic fields \citep{Bale2016}. While the sampling rate of the MAG varies between $\sim$73.24 Sa $s^{-1}$, 146.48 Sa $s^{-1}$, and 292.96 Sa $s^{-1}$ during Encounter 1, the magnetic field is sampled at a constant rate of 146.48 Sa $s^{-1}$ during Encounter 2. 
 Measurements of the thermal plasma come from the PSP/SWEAP instrument suite \citep{Kasper2016}. \cite{Huang_2020} and \cite{Halekas2020} have estimated parallel and perpendicular temperatures (with respect to the direction of magnetic field) of protons ($T_{\parallel,p}$ and $T_{\perp,p}$) and electrons ($T_{\parallel,e}$ and $T_{\perp,e}$), respectively by applying bi-Maxwellian fits to velocity distribution functions (VDFs) of protons evaluated from SWEAP/SPC \citep{Case_2020} and electrons evaluated from SWEAP/SPANe \citep{Whittlesey_2020} measurements. We use plasma parameters evaluated by averaging over SPC and SPANe observations as inputs into the \texttt{PLUME} solutions.

In SKM23, we assessed 15-minute long intervals from Encounters 1 and 2 of PSP/FIELDS and SWEAP observations, following the procedure described in Appendix A.1 of SKM23, where we obtained 2198 intervals from the available FIELDS observations; further prescreening for the availability of SWEAP observations yielded 2118 "good" intervals. For each of these intervals, we evaluated interval-averaged values of the magnetic field amplitude ($B$), proton number density ($n_p$), solar wind velocity ($\mathbf{V}_{\text{SW}}$), proton gyroradius ($\rho_p$), and Alfv\'en speed $\left( v_A = \frac{B}{\sqrt{\mu_0 n_p m_p}} \right)$ as described in appendices A.2 and C of SKM23. Here $\mu_0$ is the permeability of free space and $m_p$ is the proton mass. 

In this work, we further evaluate the parallel temperature disequilibrium ($T_{\parallel,p}/T_{\parallel,e}$), proton and electron temperature anisotropies ($T_{\perp,p}/T_{\parallel,p}$ and $T_{\perp,e}/T_{\parallel,e}$), parallel proton plasma beta $\left( \beta_{\parallel,p} = \frac{n_p  k_B T_{\parallel,p}}{(B^2 \big / 2 \mu_0)} \right)$, and parallel proton thermal velocity $ \left( v_{\text{th},\parallel,p} = \sqrt{\frac{2 k_B T_{\parallel,p}}{m_p}} \right)$. Here $k_B$ is the Boltzmann constant. 
The proton and electron temperature anisotropy estimates are discarded when their VDF measurements do not meet the data quality criteria for bi-Maxwellian fitting (see appendices of \cite{Huang_2020}), further excluding 366 intervals in Encounter 1 and 2, leaving us with 1752 total intervals for this study. 

\subsection{ Overview of SKM23: Characterizing Turbulent Energy Spectra and Estimating Heating Rates due to Landau Damping}

Figure \ref{fig:wave_turb_diss_master_plot} shows the overview of the SKM23 method combined with the current work to estimate dissipation rates of turbulence and ion-scale waves via the Landau and cyclotron resonances, respectively for an example PSP interval from 10th November 2018, 17:53:58 - 18:08:53 where ion-scale waves are observed.

In each interval analyzed in SKM23, we removed reaction wheel noise from the FIELDS magnetic field observations, downsampled the data to 10 Hz by applying an antialiasing filter, and evaluated the magnetic field wavelet energy spectrum (green dashed curve in panel (a)) using a Morlet wavelet \citep{Terrence_compo} across $f_{\text{log\_bin}}$, the median bin values obtained by splitting the DFT frequency domain of the 15 minute-\textbf{B} time series into logarithmic bins.

We then distinguished the energy spectrum of turbulence (blue in panels (a,c)) from the energy spectra of ion-scale waves, $|\tilde{B}(f)|^2_{wave\_psp}$ (black in panels (a,b)), by using a high degree of circular polarization of the latter as a criterion. Circular polarization in the plane perpendicular to the local magnetic field is quantified by spacecraft frame polarization, $\sigma$ \citep{Bowen_ICW}, which is evaluated using the wavelet-transformed magnetic field components, $W_{x,y} (s,t)$, in a scale sensitive coordinate system, $(X,Y,Z)$, where $\mathbf{Z}$ is parallel to the local magnetic field at the scale under consideration, and $\mathbf{X}$ and $\mathbf{Y}$ define the plane perpendicular to $ \mathbf{Z}$, 
\begin{equation} \label{eq:sigma_def}
 \sigma (s,t) = \dfrac{-2  \text{Im}\big[W_x(s,t)W_y^*(s,t)\big]}{|W_x(s,t)|^2 + |W_y(s,t)|^2}.
\end{equation} 
Here $s$ is the inverse frequency and $^*$ represents the complex conjugate. Using our phase and Fourier conventions, in the plasma rest frame, $\sigma$ = 1 represents ion-resonant/left-handed waves (ICWs) and $\sigma$ = -1 represents electron-resonant/right-handed waves (FMWs) (see Appendix \ref{app:polarization} for a detailed mapping between helicity ($\sigma_m$) and polarization in relevant frames). We applied a threshold of 0.7 on values of $|\sigma|$ for wave selection (see Appendix B.4. of SKM23 for justification of this criterion). Note that there is an additional minus sign in the definition of $\sigma$ (Equation \ref{eq:sigma_def}) compared to SKM23. Our previous findings are unaffected by this sign difference, as SKM23 used only the magnitude of $\sigma$ to identify and exclude coherent wave power. Additionally, we called $\sigma$ cross-coherence in SKM23, while we accurately call it spacecraft-frame polarization here. In this work, we additionally filter out the energy of turbulent fluctuations with random high values of polarization from $|\tilde{B}(f)|^2_{wave\_psp}$ (described in Appendix \ref{App:wave_identification}) and further distinguish right- (RH) and left-handed (LH) spacecraft frame polarized wave energy spectrum, $|\tilde{B}(f)|^{2^{LH/RH}}_{wave\_psp}$. 

To the turbulent energy spectra, we applied a 1-D cascade model described in \cite{Howes2008}. The model assumes a critically balanced \citep{GS1995} distribution of low-frequency, Alfv\'enic turbulence from the inertial to dissipation ranges, connecting the MHD and kinetic descriptions, and calculates the linear Landau damping rates of kinetic Alfv\'en waves onto protons and electrons (blue and orange, respectively in panel (e)) as a function of the spatial scale perpendicular to the mean magnetic field, producing a steady-state solution for a one-dimensional \cite{Batchelor1953}-like model. In intervals where the model energy spectra (red in panel (c)) fit well with the observed turbulent energy spectra (as shown in panel (c)), we assume the model to be accurate and evaluate heating rates via  Landau damping.

In this work, we use linear cyclotron damping rates of ICWs onto protons and FMWs onto electrons (blue and orange, respectively in panel (d)) to evaluate the heating rates due to the dissipation of ion-scale waves.

\begin{figure} 
    \centering
    \includegraphics[scale = 0.35]{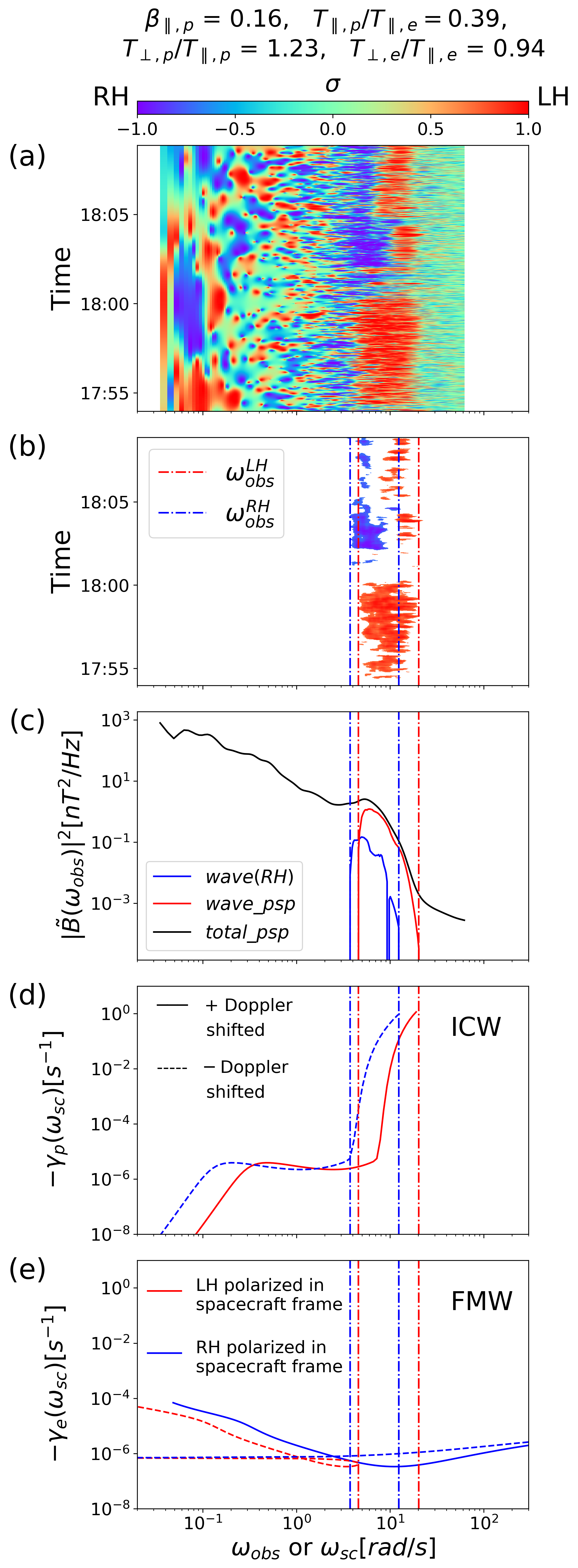}
    \caption{Overview of methodology to estimate ion-scale wave dissipation in the PSP interval (considered in Figure \ref{fig:wave_turb_diss_master_plot}) on 10th November 2018 17:53:58 - 18:08:53 where both RH and LH polarized waves are observed: (a) The spacecraft-frame polarization spectrum, $\sigma (\omega,t)$, (b) the frequency ranges where LH and RH polarized waves are observed ($\omega_{\text{obs}}^{LH/RH}$, red and blue dashed-dotted, respectively), and (c) their corresponding energy spectra (red and blue) along with the total energy spectrum (black). (d, e) The damping rates of ICWs and FMWs onto protons and electrons, respectively as functions of spacecraft-frame frequencies that are evaluated using \texttt{PLUME} for all possible Doppler shifts.}
    \label{fig:wave_diss_methodology}
\end{figure}

\subsection{Removing non-wave Signatures and Identifying Ion-scale Wave Frequencies in the Spacecraft Frame} \label{subsec_sigma_smooth}
In SKM23, we observed significant energy in ion-scale waves in the first two PSP encounters, as has been previously reported, e.g., \cite{Bowen_ICW}, \cite{Verniero_2020}, \cite{Zhao_2021}, and \cite{Liu_2023}. Panel (a) of Figure \ref{fig:wave_diss_methodology} shows $\sigma(s,t)$ for the same example PSP 15-minute interval on November 10, 2018, where ion-scale waves are observed. Persistent high values of $|\sigma|$ associated with these waves are observed over frequency bands near the proton cyclotron frequency, $f_{cp} = \frac{q_p B}{2 \pi m_p}$, and last for times varying from several seconds to a few hours. Here $q_p$ is the charge of the proton.
In the absence of ion-scale waves, non-persistent random $\sigma$ values corresponding to the background turbulent fluctuations are observed over timescales of $\sim$ 1/frequency, i.e. the time resolution of the wavelet transform. By employing a time-smoothing routine on $\sigma(s,t)$, we identify the frequency band, $f^{(RH/LH)}_{\text{obs}}$ where the ion-scale waves are observed (shown in panel (b) of Figure \ref{fig:wave_diss_methodology}) and evaluate their energy spectra, $|\tilde{B}(f)|^{2^{LH/RH}}_{wave\_psp}$ (shown in panel (c) of Figure \ref{fig:wave_diss_methodology}) as described in Appendix \ref{App:wave_identification}. Here $\omega^{(RH/LH)}_{\text{obs}} = 2\pi f^{(RH/LH)}_{\text{obs}} $ is the angular frequency band where the waves are observed.

\subsection{Calculating Damping Rates at $\omega_{\text{obs}}$} 
\label{sec:Doppler_shifts}
The four MHD modes - the Alfv\'en, fast and slow magnetosonic waves, and the entropy mode have kinetic counterparts in a collisionless plasma that can be relatively lightly damped (see \cite{Klein_PhD_thesis} for an overview). 
The assumption of symmetric VDFs produces degenerate solutions for the three propagating modes, and therefore the damping rates of the forward and backward solutions are identical.
The impact of the mode propagation direction is considered in the calculation of the spacecraft to plasma frame conversion, see Eqn.~\ref{eqn:om_sc_1}. At small parallel scales, Alfv\'en waves extend to ICWs and FMWs extend to whistler waves. 
The \texttt{PLUME} dispersion solver \citep{Klein_Howes_2015_PLUME} calculates the hot linear plasma dispersion relation $|D(\omega, \gamma; \mathbf{k}, \mathcal{P})| = 0$ for bi-Maxwellian VDFs of ions and electrons in a uniform magnetized environment. 
The solutions of the dispersion relation yield dimensionless angular frequencies and damping rates $(\omega/\Omega_p, \gamma/\Omega_p)_{\texttt{PLUME}}$ in the plasma frame--which are normalized to physical units via the observed proton gyrofrequency, $\Omega_p = 2 \pi f_{cp}$--of different plasma modes as a function of wavevector, ($k_\perp$, $k_\parallel$) and equilibrium plasma parameters $\mathcal{P}$.

For this work, we consider the parameter space $\mathcal{P} = \left[ \beta_{\parallel,p}, \frac{T_{\parallel,p}}{T_{\parallel,e}}, \frac{T_{\perp,p}}{T_{\parallel,p}}, \frac{T_{\perp,e}}{T_{\parallel,e}}, \frac{v_{\text{th},\parallel,p}}{c} \right]$, representing a proton-electron plasma with temperature anisotropies for both species. 
Additionally, the eigenfluctuations of the distribution function and the electromagnetic fields corresponding to the modes are evaluated. 
For each mode, using the perpendicular electric field eigenfunctions and components of proton and electron susceptibility tensor, the damping rates onto the species normalized to frequency, $(\gamma_{p,e}/\omega)_{PLUME}$ are evaluated, following the prescription described in \cite{Quataert:1998}. Values of $\gamma_{PLUME}$ are negative for a damped wave, and positive when a wave is linearly unstable, which can arise in this work over narrow bands of $k_\parallel$ driven by proton temperature anisotropy (see Sections \ref{subsec:wave_regionality} and \ref{sec:discussion}).

\begin{figure*}
    \centering  
    \plotone{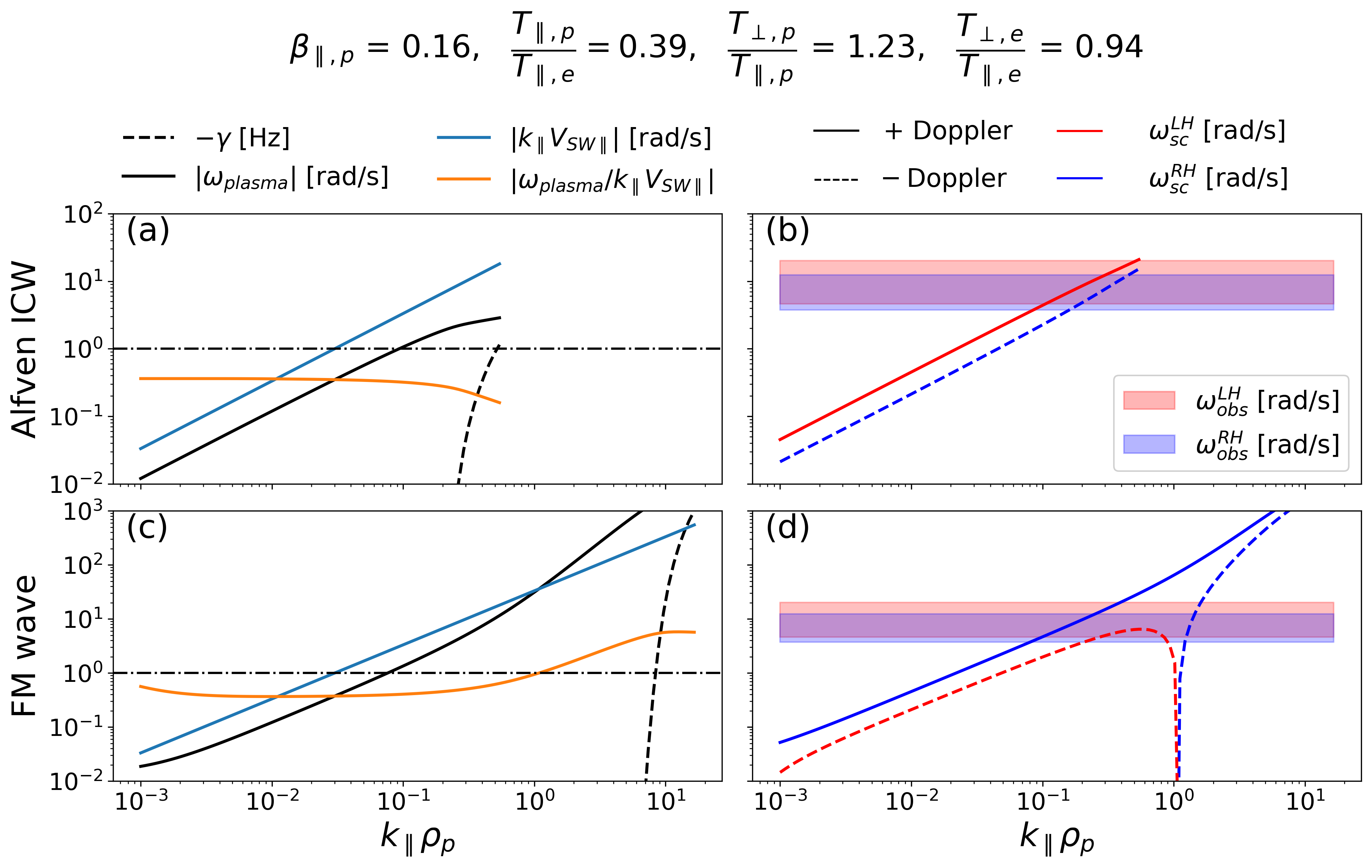}
    \caption{All possible Doppler shift cases for an ICW (panels a,b) and FMW (panels c,d) for the PSP interval considered in Figure \ref{fig:wave_diss_methodology} where both RH and LH polarized waves are observed. The behavior of Doppler-shifted ICWs and FMWs depends on their intrinsic direction of propagation, \textbf{k}, with respect to $\mathbf{V}_{\text{SW} \parallel}$, the component of $\mathbf{V}_{\text{SW}}$ parallel to \textbf{B} and the relative magnitudes of plasma frame frequencies, $\omega_{\text{plasma}}$, and $k_\parallel V_{\text{SW}\parallel}$. Each Doppler shift case is described in the text of section \ref{sec:Doppler_shifts}. Here \texttt{PLUME} dispersion solutions where waves are heavily damped, $\gamma/\omega \ge e^{-1} $ are discarded, where $e = 2.718$.}

    \label{fig:all_Doppler_cases}
\end{figure*}

For each interval, we evaluate the values of $\mathcal{P}$ using PSP observations, input them into \texttt{PLUME}, identify the four least damped modes at MHD scales ($k_\perp \rho_p$  = $k_\parallel \rho_p$ = $10^{-3}$), follow their dispersion solutions to smaller parallel scales up to $k_\parallel \rho_p = 10^2$ with $k_\perp \rho_p$ held constant. Left panels of Figure \ref{fig:all_Doppler_cases} show the plasma-frame angular frequencies (black-solid), $\omega_{\text{plasma}} = 2\pi f_{\text{plasma}}$, and damping rates (black-dashed) of the ICW (panel (a)) and FMWs (panel (c)) that are evaluated using \texttt{PLUME} for the example interval considered in Figure \ref{fig:wave_diss_methodology}. For the observed values of $\mathcal{P}$ in Encounters 1 and 2, we accurately identify the forward and backward ICWs and FMWs and distinguish between the electron- and ion-resonant modes via the procedure described in Appendix \ref{app:mode_identify}.

The plasma frame frequencies are then Doppler shifted to the spacecraft frame via
\begin{equation}
   \omega_{sc} = \omega_{\text{plasma}} + \mathbf{k} \cdot \mathbf{V_{\text{SW}}}, 
   \label{eqn:om_sc_1}
\end{equation}
where $\omega_{sc}$ is the spacecraft-frame frequency. Here we adopt the convention that the plasma-frame wave frequencies, $\omega_{\text{plasma}}$, are positive definite, and their direction of propagation is determined by wavevector \textbf{k}. We assume that the observed ion-scale waves propagate along the magnetic field. The validity of this assumption is underlined by the first, median, and third quartiles of the distribution of $\theta_{\mathbf{k}\mathbf{B}}$, the acute angle between the wavevector and the mean magnetic field, for all intervals where ion-scale waves are observed being $5.2^\circ$, $9.7^\circ$, and $18.8^\circ$, respectively. Here, the wavevector direction is estimated by calculating the direction of minimum variance of \textbf{B} \citep{Sonnerup_1967_MVA}. Thus we rewrite Equation ~\ref{eqn:om_sc_1} as
\begin{equation}
   \omega_{sc} = \omega_{\text{plasma}} \pm k_{\parallel} \hat{\mathbf{b}} \cdot \mathbf{V_{\text{SW}}}  , \hspace{1cm}  \hat{\mathbf{b}} = \frac{\mathbf{B}}{|\mathbf{B}|}.
\end{equation} 
Here, plus (minus) sign of the $k_{\parallel} \hat{\mathbf{b}} \cdot \mathbf{V_{\text{SW}}}$ term corresponds to waves propagating along (opposite to) \textbf{B}. The sign of the $k_{\parallel} \hat{\mathbf{b}} \cdot \mathbf{V_{\text{SW}}}$ term is further determined by the angle between \textbf{B} and $\mathbf{V}_{\text{SW}}$. The sign of the $k_{\parallel} \hat{\mathbf{b}} \cdot \mathbf{V_{\text{SW}}}$ term is thus determined by the direction of wave propagation with respect to the parallel component of $\mathbf{V}_{\text{SW}}$, $\mathbf{V}_{\text{SW} \parallel} = \hat{\mathbf{b}} \cdot \mathbf{V}_{\text{SW}}$.    
For a wave propagating along (opposite to) the $\mathbf{V_{\text{SW} \parallel}}$  direction i.e. outward (inward) with respect to the Sun, the sign of $k_{\parallel} \hat{\mathbf{b}} \cdot \mathbf{V}_{\text{SW}}$ term is positive (negative) and thus the wave is positive (negative) Doppler shifted,
\begin{equation} \label{ds_eq}
\omega_{sc} = \omega_{\text{plasma}} \pm k_{\parallel}  V_{\text{SW} \parallel} .
\end{equation}

 Consequently, depending on the relative magnitudes of the $\omega_{\text{plasma}}$ and  $k_{\parallel}  V_{\text{SW} \parallel}$ terms on the R.H.S of Eqn.~\ref{ds_eq}, the sign of $\omega_{sc}$ is either positive or negative. A positive (negative) $\omega_{sc}$ value corresponds to retaining (reversing) the plasma-frame direction of propagation and polarization of the wave when Doppler shifted to the spacecraft frame. Figure \ref{fig:all_Doppler_cases} shows the possible Doppler shift behaviors of ICWs (panels (a,b)) and FMWs (panels (c,d) for the example PSP interval considered in Figure \ref{fig:wave_diss_methodology}. The magnitudes of the $\omega_{\text{plasma}}$ (black-solid) and $k_{\parallel}  V_{\text{SW} \parallel}$ (light blue) terms, and their ratio (orange) are plotted as functions of parallel scale in panels (a,c). Further, the resultant spacecraft frame frequencies with LH ($\omega_{\text{sc}}^{LH}$, red) and RH ($\omega_{\text{sc}}^{RH}$, blue) polarizations upon positive (solid) and negative (dashed) Doppler shifts are plotted as functions of parallel scale in panels (b,d).
 For the ICW and FMWs at a particular parallel scale, $k_\parallel$, there are three possible outcomes due to Doppler shifting; 
\begin{enumerate}
    \item An ICW (FMW) propagating along $\mathbf{V_{\text{SW} \parallel}}$ is positively Doppler shifted by the solar wind, retaining its plasma frame polarization and is observed LH (RH) polarized in spacecraft frame.
    \item An ICW (FMW) propagating opposite to $\mathbf{V_{\text{SW} \parallel}}$ is negatively Doppler shifted by solar wind and if: 
    \begin{enumerate}
        \item $|\omega_{\text{plasma}}| > k_{\parallel}  V_{\text{SW} \parallel}$, it retains its plasma frame polarization and is observed LH (RH) polarized in spacecraft frame.
        \item $|\omega_{\text{plasma}}| < k_{\parallel}  V_{\text{SW} \parallel}$, it reverses its plasma frame polarization and is observed RH (LH) polarized in spacecraft frame.
    \end{enumerate}    
\end{enumerate}
These behaviors are depicted in the right panels of Figure \ref{fig:all_Doppler_cases} and summarized in Table~\ref{tab:doppler}.

\begin{table}
\setlength{\tabcolsep}{8pt}
    \centering
    \begin{tabular}{m{1.75cm}|c c c}
    \hline
    \multirow{4}{1cm}{Plasma frame polarization} & \multicolumn{3}{c}{  Spacecraft frame polarization}\\ [1 ex] \cline{2-4}
    & \multirow{3}{1.5cm}{+ Doppler shift} & \multicolumn{2}{|c}{- Doppler shift}
    \\ [1 ex] \cline{3-4}
   & & \multicolumn{1}{|c}{$|\omega_{\text{plasma}}| >$} & \multicolumn{1}{|c}{$|\omega_{\text{plasma}}| < $}  \\
   & & \multicolumn{1}{|c}{$ k_{\parallel}  V_{\text{SW} \parallel}$} & \multicolumn{1}{|c}{$ k_{\parallel}  V_{\text{SW} \parallel}$} \\[1.5 ex]
    \hline 
    ICW (LH) & LH & \cellcolor{black!15} LH & RH \\
    
    FMW (RH) & RH & RH & $\text{LH}^{*} $ \\
    \hline
    \end{tabular}
    \caption{Three possible Doppler shifted polarizations for ICWs and FMWs. The shaded Doppler shift doesn't occur in Encounters 1 and 2. $^{*}$ Negative Doppler shifted FMW frequencies overlap with observed frequencies over two distinct bands which are represented with a suffix $\_1,2$.}
    \label{tab:doppler}
\end{table}

\begin{figure}[hbt!]
    \centering
    \includegraphics[scale = 0.35]{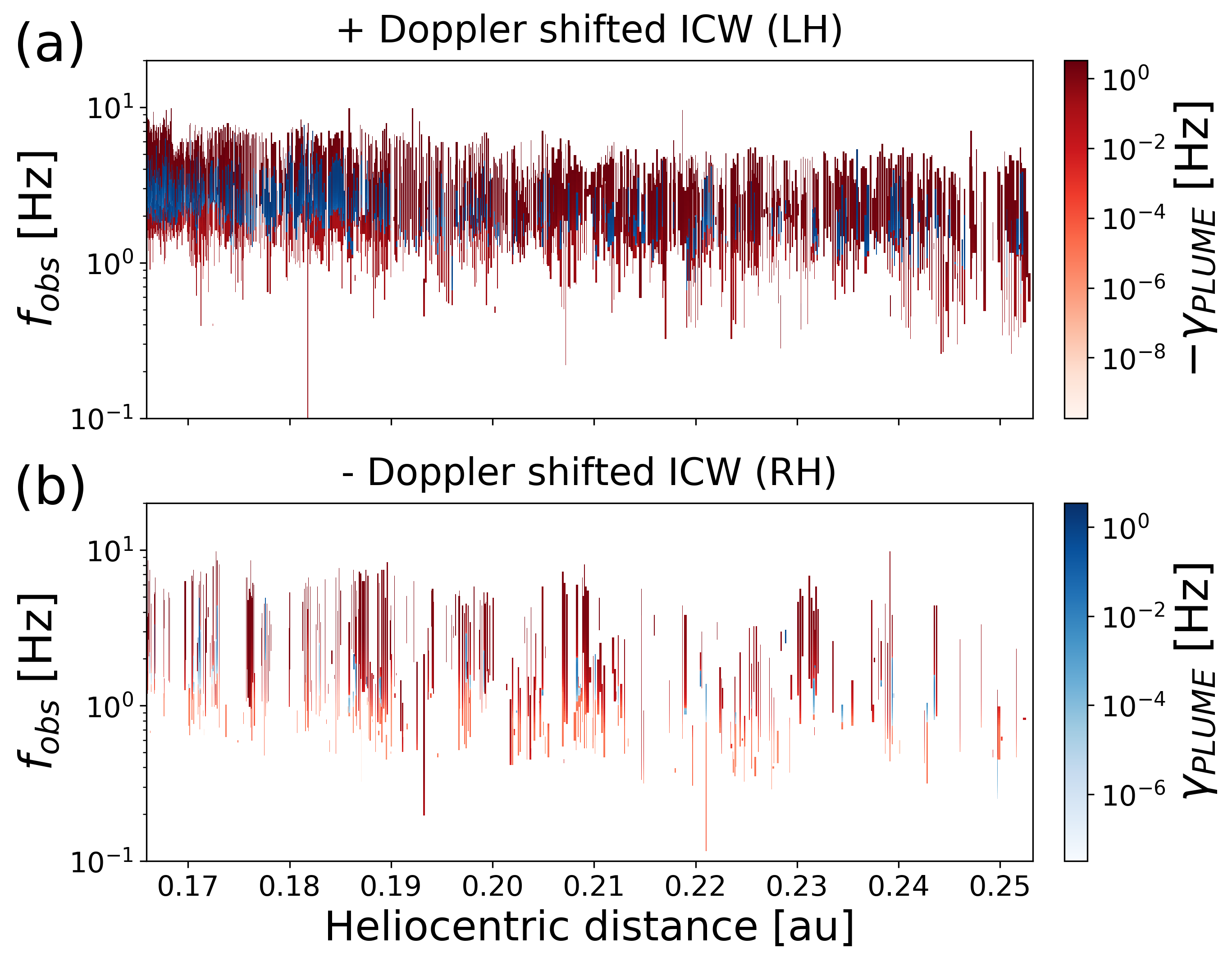}
    \caption{\revise{Contour maps of \texttt{PLUME} damping (red) and growth (blue) rates as a function of heliocentric distance and $f_{obs}$, the frequencies where ion-scale waves are observed by PSP during Encounters 1 and 2 for the positive and negative Doppler-shifted ICW cases.}}
    \label{fig:gamma_omega_spectrogram}
\end{figure}

For ICWs (panels (a,b)), $|\omega_{\text{plasma}}|$ is usually less than $k_{\parallel}  V_{\text{SW} \parallel}$ at all $k_\parallel$ (see orange curve), and they have LH (RH) polarization upon positive (negative) Doppler shift which is plotted as red-solid (blue-dashed) curve. 
However, for FMWs (panels (c,d)), $|\omega_{\text{plasma}}|$ become strongly dispersive at small parallel scales gaining larger phase velocities, allowing them to exceed $k_{\parallel}  V_{\text{SW} \parallel}$ in magnitude (see orange curve). While a positive Doppler-shifted FMW has RH polarization (blue-solid curve), a negative Doppler-shifted FMW can switch polarizations from LH (red-dashed curve) to RH (blue-dashed) moving from smaller to larger $k_\parallel$. 
Consequently, for these waves, the magnitude of $\omega_{sc}$ increases with $k_\parallel$, reaches a maximum, and then decreases before reversing sign (see the red-dashed curve). The Doppler shift routine employed in this work and the behaviors of ICWs and FMWs upon Doppler shifting are consistent with \cite{Bowen_Outward_ICW_2020}, \cite{Verniero_2020}, and \cite{Mcmanus_2024}. 

In an interval where an ion-scale wave with a RH or LH polarization is observed such as the one depicted in panel (a) of Figure \ref{fig:wave_diss_methodology} and panels (b,d) of Figure \ref{fig:all_Doppler_cases}, its occurrence can be ascribed to the three discussed Doppler shift cases each for either polarization. We check for overlaps (as seen in panels (b,d) of Figure \ref{fig:all_Doppler_cases}) of $\omega_{\text{obs}}$ (blue/red shaded region) with $\omega_{sc}$ (blue/red, solid and dashed curves) for all the Doppler shift cases and consider damping rates corresponding to the overlapping $\omega_{sc}$. For a negative Doppler shifted FMW (like the red dashed curve in panel (d) of Figure \ref{fig:all_Doppler_cases}), there could be two bands of $\omega_{sc}$ overlapping with $\omega_{\text{obs}}$ which are named with suffixes $_1,_2$ in the rest of this paper (e.g. Figure \ref{fig:wave_local_generation}). To estimate energy dissipation rates in such cases, we identify the band of $\omega_{sc}$ that encompasses or is nearest to $k_\parallel \rho_p = 0.3$ as the relevant region and use the corresponding damping rates. This selection criterion is justified as the most likely wavevector region that supports the emission of FMWs from the parallel firehose instability \citep{Gary_SP_1993, verscharen2019} is typically around $ k_\parallel \rho_p \sim 0.3$. The \texttt{PLUME} damping rates, $\gamma_{p,e\_PLUME} (\omega_{sc})$ are then linearly interpolated at $\omega_{\text{obs}}$ when the overlaps occur. Because heavily damped waves cannot propagate, \texttt{PLUME} dispersion solutions with $\omega \le e \times \gamma$ are discarded, where $e = 2.718$. As a result of this selection, occasionally, $\gamma_{p,e\_PLUME}$ is unavailable at higher frequencies of $\omega_{\text{obs}}$. We extend the higher end value of $\gamma_{p,e\_PLUME}$ to these frequencies. Panels (d) and (e) of Figure \ref{fig:wave_diss_methodology} show $\gamma_{p,e\_PLUME} (\omega_{\text{obs}})$ corresponding to all Doppler shift cases where the overlaps occur. 

Figure \ref{fig:gamma_omega_spectrogram} shows $\gamma_{p,e\_PLUME}$ as a function of $f_{\text{obs}}$, the frequency bands where ion-scale waves are observed in Encounters 1 and 2 and heliocentric distance for the positive and negative Doppler-shifted ICW cases. The $\gamma_{p,e\_PLUME}$ values corresponding to all the FMW Doppler shift cases are 6 orders of magnitude lower than the $\gamma_{p,e\_PLUME}$ values shown in Figure \ref{fig:gamma_omega_spectrogram}. The insignificance of FMW cyclotron damping rates is consistent with the resonance between FMW's rotating electric field and gyrating electrons being negligible at proton scales where the waves are observed and are assumed to dissipate. We see a considerable fraction of intervals with unstable waves, where growth rates are comparable to damping rates. However, the supported region of unstable modes occurs over a narrower frequency band compared to $f_{obs}$, which can lead to a net damping of cyclotron waves rather than emission, as discussed in Sec.~\ref{sec:discussion}.

\subsection{Evaluating Energy Dissipation Rates onto Protons and Electrons}

We transform the observed solar wind magnetic field fluctuations to velocity units ([m/s]) and evaluate the energy spectra of observed ion-scale waves in SI units as

\begin{equation}
 |\tilde{B}(f)|_{wave\_psp}^2[\text{m}^2 \text{s}^{-2} \text{Hz}^{-1}] =  \frac{|\tilde{B}(f)|_{wave\_psp}^2[\text{T}^2 \text{Hz}^{-1}]}{\mu_0 m_p n_p[\text{m}^{-3}]}.
\end{equation}

In each interval where ion-scale waves are observed, the dissipation rates are estimated as the sum over the product of $\gamma_{p,e}$ and the magnetic power spectrum over the identified frequency range
\begin{equation}
\begin{split}
    Q_{p,e} [\text{m}^2 \text{s}^{-3}] = & - \sum \left(|\tilde{B}(\omega_{\text{obs}})|^{2^{RH/LH}}_{wave\_psp}[\text{m}^2 \text{s}^{-2} \text{Hz}^{-1}] \times \right. \\
   & \left. \Delta f(\omega_{\text{obs}}) [\text{Hz}] \times \gamma_{p,e\_PLUME}(\omega_{\text{obs}})[\text{Hz}] \vphantom{|^2_{wave\_psp}} \right), 
    \end{split}
\end{equation}
where $\Delta f$ is the array of widths of logarithmic frequency bins at which energy spectra are evaluated, and $Q_{p,e}$ is in units of energy rate per unit mass. This method of estimating heating rates is similar to the method employed in SKM23.

\section{Occurrence of ion-scale waves in the young solar wind} 
\label{sec:occurence}

\begin{figure}[hbt!]
    \centering
    \includegraphics[scale = 0.3]{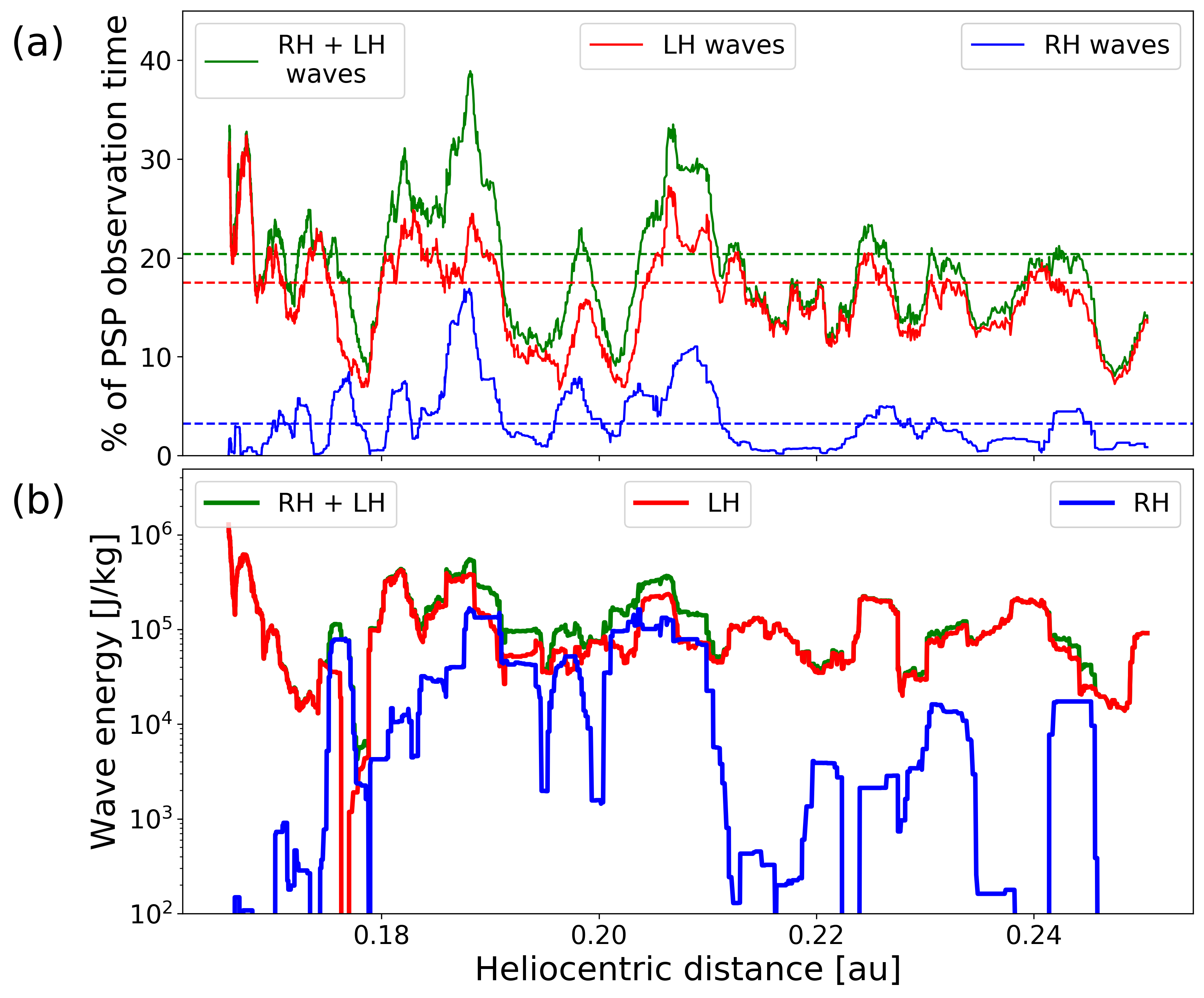}
    \caption{The moving averages of: panel (a) the fraction of observation time during which LH (red), RH (blue) polarized waves, and either/both of them (green) are shown as a function of heliocentric distance along with their means over all E1 and E2 intervals (dashed). Panel (b) the energy in LH (red), RH (blue) polarized waves, and their sum (green) are plotted as a function of heliocentric distance.}
    \label{fig:RH vs LH statistics}
\end{figure}

 In all intervals where ion-scale waves are observed, we estimate their duration, $T_{wave}$, by evaluating the total time length of the $(f-t)_{coherent}$ domain, the region where a high spacecraft frame polarization of $|\sigma| \ge 0.7$ is observed at one frequency at least in $f_{\text{log}\_\text{bin}}$ after time averaging the spacecraft frame polarization spectrum, $\sigma(s,t)$ as described in subsection \ref{subsec_sigma_smooth}. 
 We then evaluate the percentage of interval observational time (i.e. $\sim$ 15 minutes) during which waves are observed (their moving averages shown in panel (a) of Figure \ref{fig:RH vs LH statistics}). 
 We find that persistent ion-scale waves are present during 20.37$\%$ of Encounters 1 and 2 observations, of which LH polarized (in the spacecraft frame) waves are dominant and are observed during 85.86$\%$ of the time (17.49$\%$ of total observational time), and RH waves are observed during 15.8$\%$ of the time (3.22$\%$ of total observational time). Both the RH and LH waves are observed simultaneously during 0.34\% of total observational time. The dominance of LH waves is consistent with \cite{Bowen_ICW} and \cite{Liu_2023}. We further examine the dependence of wave detection on the angle between \textbf{B} and $\mathbf{V}_{\text{SW}}$ by subdividing Encounters 1 and 2 magnetic field observations into $\sim$0.873 second intervals (corresponding to the cadence of SPC measurements) and evaluating the acute angle between \textbf{B} and $\mathbf{V}_{\text{SW}}$, $\theta_{B V_{\text{SW}}}$. We find that the first, median, and third quartiles of the $\theta_{B V_{\text{SW}}}$ distribution in intervals with waves are 17.3, 25.4, and 34.4, respectively, demonstrating that (consistent with \cite{Bowen_ICW} and \cite{Woodham_2021}), ion-scale waves are preferentially observed when \textbf{B} and $\mathbf{V}_{\text{SW}}$ are aligned. When $\theta_{B V_{\text{SW}}}$ is oblique, the wave occurrence is undetermined and the wave signatures could be dominated by anisotropic turbulence in which case the percentages reported above would be a lower bound on the ion-scale wave occurrences.  
 
 In all intervals, we further evaluate the energy in ion-scale waves, $E_{\text{wave}}^{RH/LH} = \sum \left(|\tilde{B}(\omega_{\text{obs}})|^{2^{RH/LH}}_{wave\_psp}
\Delta f(\omega_{\text{obs}}) \right)$ (shown in panel (b) of Figure \ref{fig:RH vs LH statistics}). We find that LH (RH) polarized waves constitute 90.05$\%$ (9.95$\%$) of the total energy in ion-scale waves in Encounters 1 and 2 observations.

Note that ion-scale waves are observed during a higher fraction i.e. 29.27\% of observational time in intervals that were discarded due to the unavailability of SWEAP observations or temperature anisotropy estimations. However, the statistics of ion-scale wave duration and energy remain similar to Figure \ref{fig:RH vs LH statistics} upon inclusion of the discarded intervals.

\subsection{Observed ion-scale waves are consistent with transient ICWs} \label{subsec:wave_regionality}

\begin{figure*}[hbt!]
    \centering
    \plotone{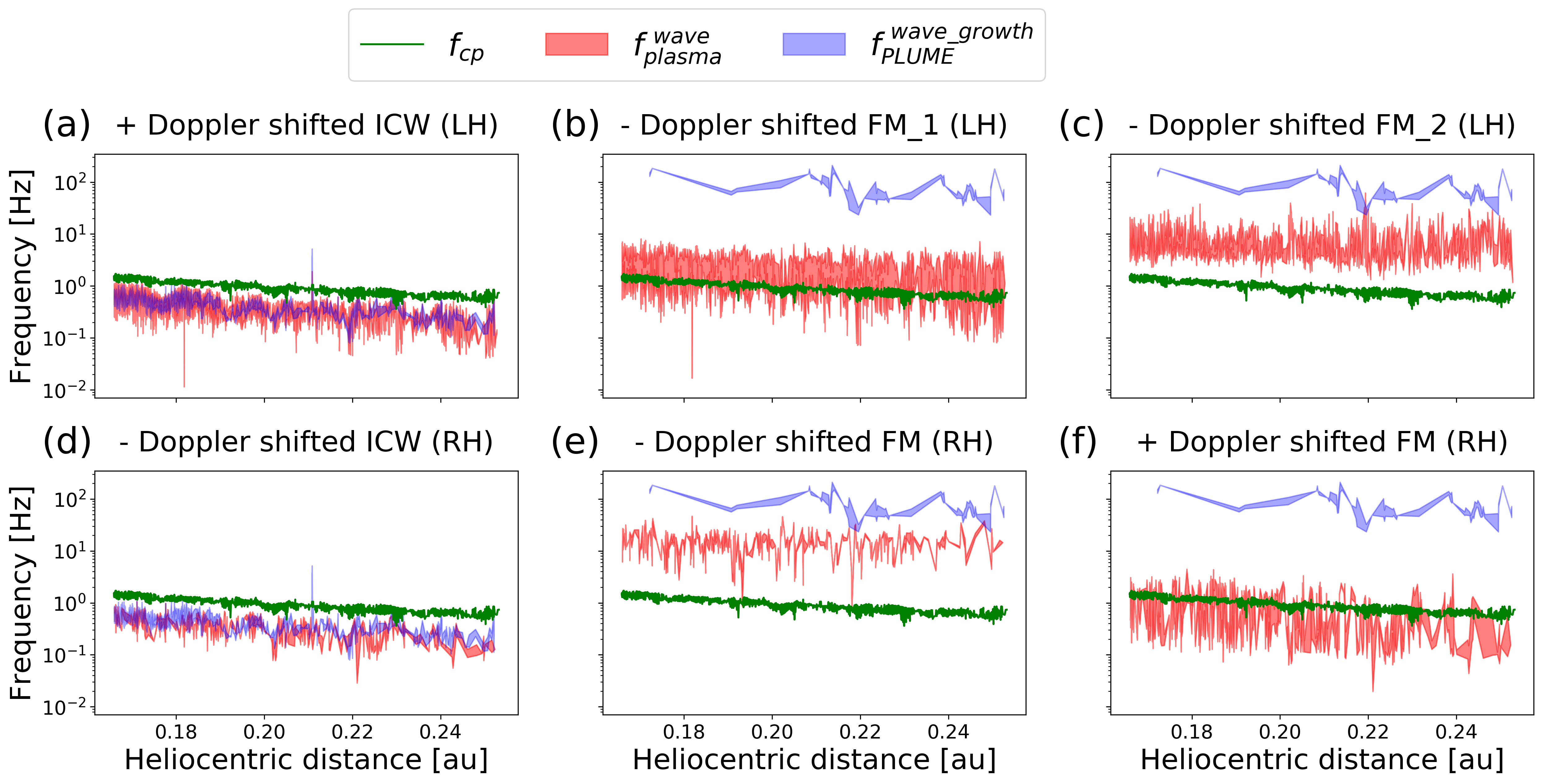}
    \caption{\revise{The plasma-frame frequency ranges corresponding to the spacecraft frame frequencies at which waves are observed ($f_{\text{plasma}}^{\text{wave}}$, red shaded), the frequencies in the plasma frame where ICWs/FMWs grow due to proton temperature anisotropy instabilities ($f_{PLUME}^{wave\_growth}$, blue shaded), and the local proton cyclotron frequency ($f_{cp}$, green dotted) as functions of heliocentric distance are shown.}}
    \label{fig:wave_local_generation}
\end{figure*}

\begin{figure}[hbt!]
    \centering
    \includegraphics[scale = 0.385]{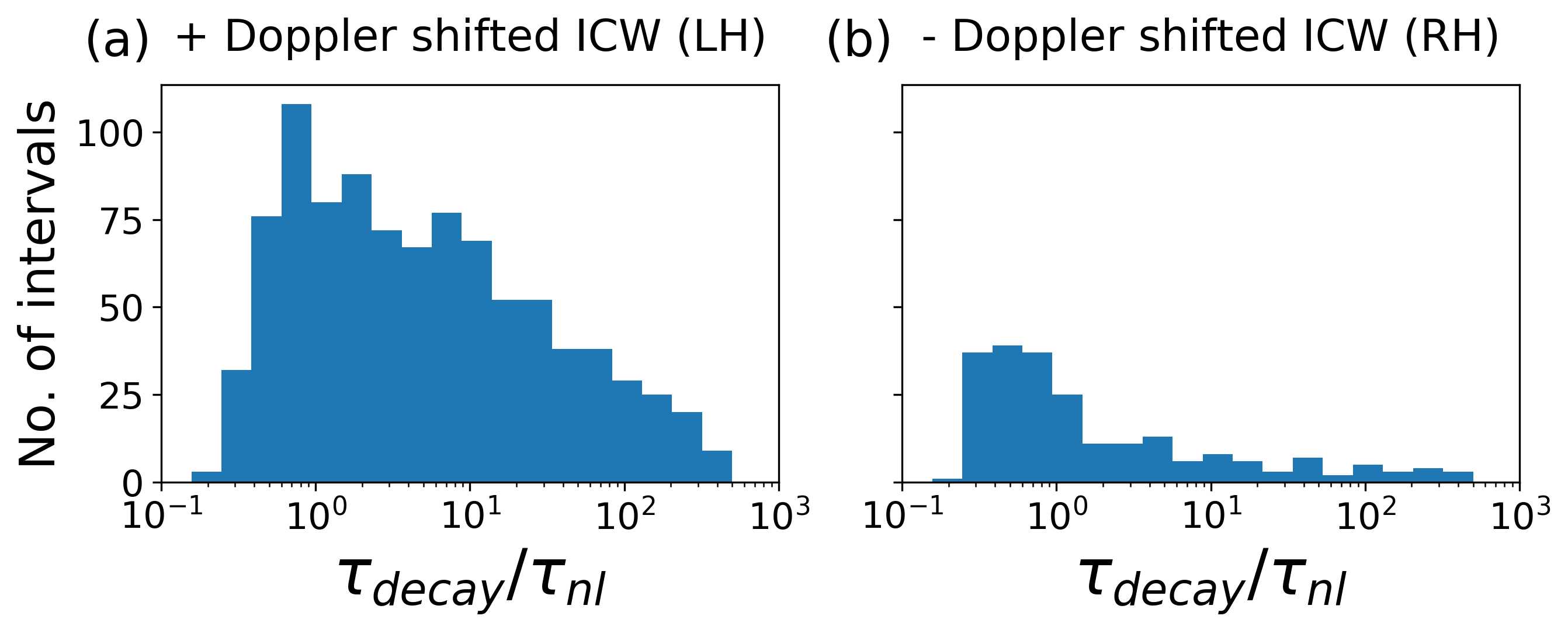}
    \caption{\revise{The ratio of the decay time, $\tau_{\text{decay}}$ to the nonlinear turbulent timescale at proton scale, $\tau_{nl}$ corresponding to the positively and negatively Doppler-shifted ICW cases.}}
    \label{fig:tau_decay_tau_nl_ratio}
\end{figure}

Using \texttt{PLUME} solutions ($f_{PLUME} = \frac{\omega_{PLUME}}{2 \pi}$) we evaluated in all the intervals where ion-scale waves are observed, the plasma-frame frequency ranges corresponding to $f_{\text{obs}}^{RH/LH}$, the spacecraft frame frequencies at which the waves are observed for all Doppler shift cases, 
\begin{equation}
\begin{split}
    f_{\text{plasma}}^{\text{wave}} = & \left[\text{min}(f_{\text{plasma}}^{\text{wave},RH} \cup f_{\text{plasma}}^{\text{wave},LH}) \right.,\\
   & \left. \text{max}(f_{\text{plasma}}^{\text{wave},RH} \cup f_{\text{plasma}}^{\text{wave},LH}) \right].
\end{split}
\end{equation}

Temperature anisotropy instabilities \citep{Gary_SP_1993} play a major role in constraining the temperature anisotropies of solar wind plasma species \citep{Hellinger_2006,Bale_2009_PRL_T_anisotropy_instabilities, Huang_2020}. The ion cyclotron (occurs when $\frac{T_{\perp,p}}{T_{\parallel,p}} > 1$) and parallel firehose (occurs when $\frac{T_{\perp,p}}{T_{\parallel,p}} < 1$) instabilities generate ICWs and FMWs, respectively.   We find that local (i.e. 15-minute interval-averaged) plasma parameters satisfy the linear instability criteria in 38\% (3.3\%) of the analyzed intervals for the ion cyclotron (parallel firehose) instability. 
In intervals where $\gamma_{PLUME}>0$, we evaluated $f_{PLUME}^{wave\_growth}$, the frequencies in the plasma frame where ICWs/FMWs are unstable and grow. Figure \ref{fig:wave_local_generation} shows $f_{\text{plasma}}^{\text{wave}}$ (red-shaded), $f_{PLUME}^{wave\_growth}$ (blue-shaded), and the local $f_{cp}$ (green) as functions of heliocentric distance. For ICWs (panels (a) and (d)), $f_{\text{plasma}}^{\text{wave}}$ overlaps well with $f_{PLUME}^{wave\_growth}$ and both the frequency ranges scale with the local $f_{cp}$. However, for FMWs (panels (b), (c), (e), and (f)),  $f_{\text{plasma}}^{\text{wave}}$ does not overlap with $f_{PLUME}^{wave\_growth}$ band that is determined by either local plasma parameter values or by the values of plasma parameters elsewhere in Encounters 1 and 2. This analysis suggests that the observed ion-scale waves are locally generated ICWs, consistent with \cite{Bowen_Outward_ICW_2020}. 
\cite{Bowen_Outward_ICW_2020} have found spacecraft-frame RH waves to be consistent with outward propagating FMWs. 
These FMWs could be driven by a proton beam instability, which we expect to be driven at frequencies consistent with the observed waves.
The effects of secondary populations will be considered in future work.

In all intervals where ion-scale waves are observed, we evaluated $\gamma^{\text{max}\_\text{wave}}$, the damping rate corresponding to the frequency in $f_{obs}$ at which the fraction of wave energy is maximum. In all intervals where  $\gamma^{\text{max}\_\text{wave}}<0$, i.e., the observed waves damp at the frequency where their energy is maximum, we estimated the timescales over which the waves decay by evaluating $\tau_{\text{decay}} = \frac{1}{-\gamma^{\text{max}\_\text{wave}}}$. We then evaluated the turbulent non-linear timescale $\tau_{nl} = \frac{k_\perp^{-2/3} k_0^{-1/3}}{v_A}$ at proton scales ($k_\perp \rho_p \sim 1$) by assuming critical balance (\cite{GS1995}). Here $ k_0 = \frac{2\pi f_{\text{break}}}{V_{\text{SW}}}$ is the outer scale evaluated by subdividing Encounter 1 and 2 observations into 6-hour intervals and then calculating $f_{\text{break}}$, the spectral break frequency at which the transition from injection to inertial scales occurs. Figure \ref{fig:tau_decay_tau_nl_ratio}
 shows the ratio $\frac{\tau_{\text{decay}}}{\tau_{nl}}$ for all the positively and negatively Doppler shifted ICW cases. For ICWs $\tau_{\text{decay}}$ has a median value of \revise{7.6} seconds and is comparable to $\tau_{nl}$. We further estimated the distance, $l_{prop} = \tau_{\text{decay}} v_{\text{phase}}^{\text{max}\_\text{wave}}$ that the observed waves propagate during the time their amplitudes damp by a factor of e = 2.718. Here $v_{\text{phase}}^{\text{max}\_\text{wave}}$ is the phase velocity at the frequency where the wave energy fraction is maximum. We find that the median value of $l_{prop}/\rho_p$ is $\sim$\revise{60}. Hence these ICWs are local and very short-lived. 

\begin{figure}[H]
    \centering
    \includegraphics[scale = 0.35]{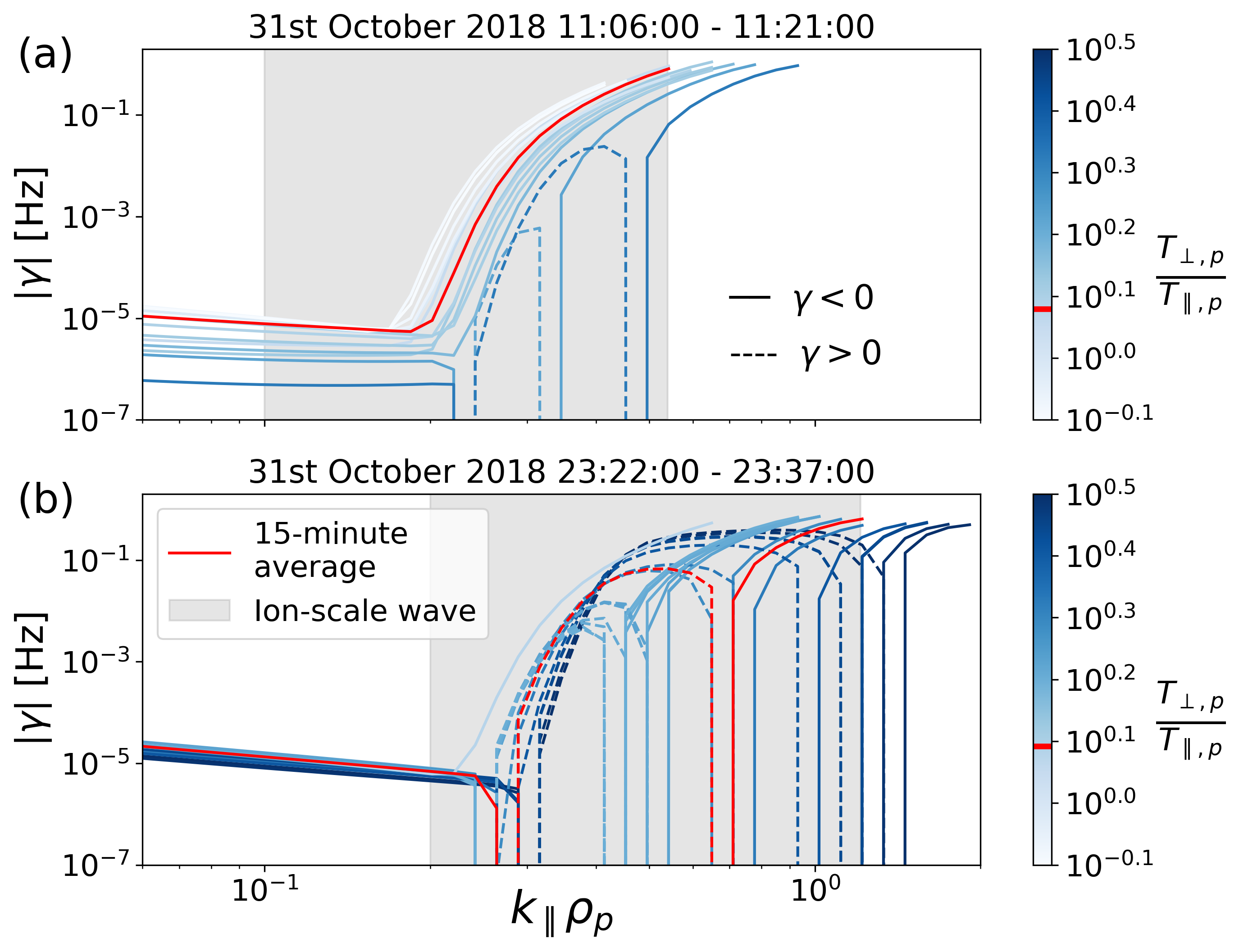}
    \caption{The \texttt{PLUME} damping (solid)/growth (dashed) rate as functions of parallel wavevector corresponding to $(\beta_{\parallel,p},\frac{T_{\perp,p}}{T_{\parallel,p}})$ values of 1-minute intervals (shades of blue) along with  the damping/growth rates corresponding to the 15-minute averaged plasma parameters (red) for two intervals where the average parameters are stable (a) or unstable (b).}
    \label{fig:T_aniso_gamma_var}
\end{figure}
 
 These waves are short-lived compared to both the cadence of the temperature anisotropy estimations used in this work \citep[1 minute estimations from][]{Huang_2020} and the time length of the intervals considered in this analysis (15 minutes), and could be driven unstable by short intervals with temperature anisotropy beyond the appropriate threshold. 
  The width of the observed wave frequencies in an interval could be a result of the aggregation of distinct waves from different regions with different plasma parameters and solar wind velocities--and hence Doppler shifts-- compared to their interval-averaged values. 
  The observed transient ICWs can be intermittently generated in unstable regions \citep{Qudsi_2020} and damp at nearby stable regions, mediating the local spatial transport of free energy associated with temperature anisotropy.
To demonstrate this, we illustrate in Fig.~\ref{fig:T_aniso_gamma_var} the ICW growth and damping rates calculated using both the 15 minute average plasma parameters compared against rates determined from the 1-minute values.
For both intervals, the one-minute averages demonstrate some stable and unstable behavior, even though the 15 minute averages are either stable or unstable.
Across all wavevectors, we see the damping rates for the stable waves determined by the 15-minute average are qualitatively similar to the 1-minute values, suggesting that the assumption that the waves damp at rates associated with the plasma parameters from the 15 minute averages is a reasonable statistical estimation.
Therefore, we argue that the estimates of cyclotron damping rates in stable regions as well as the estimates of $f_{PLUME}^{wave\_growth}$, $\tau_{\text{decay}}$, and $l_{prop}$ that are evaluated using interval averaged 
temperature anisotropy values are reasonable.

 Higher cadence temperature anisotropy observations (e.g. at the 7s cadence in \cite{Verniero_2020} and \cite{Mcmanus_2024}) could be used to better analyze the correlation between observed temperature anisotropies and waves and to determine the specific instabilities that could drive the observed waves and obtain a more precise, localized measurement of the wave-particle interaction.

\section{Significant ICW dissipation onto young solar wind protons }
\label{sec:dissipation}

\begin{figure*}[hbt!]
    \centering
    \includegraphics[scale = 0.385]{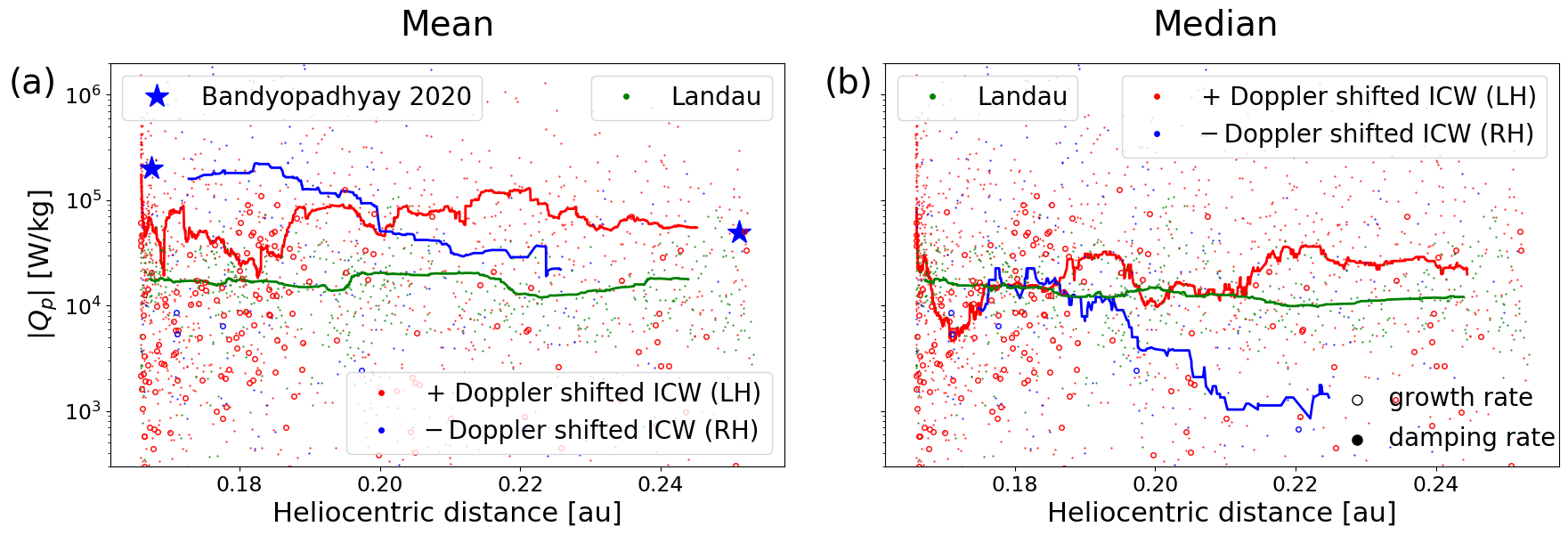}
    \caption{\revise{Estimated dissipation rates of observed ion-scale waves via cyclotron damping (this work) and turbulence via Landau damping (SKM23) in all the analyzed intervals from PSP Encounters 1 and 2. The dissipation rates of positively (red) and negatively (blue) Doppler shifted ICWs onto protons, the turbulent dissipation rates onto protons via Landau damping (green), and the inertial scale energy cascade rates evaluated using PSP observations by \cite{Bandyopadhyay_2020} (blue star) are shown.  Panel (a) shows moving means of all $Q_p$ values and Panel (b) shows moving medians of net damped intervals $Q_p<0$. Here dots (circles) represent magnitudes of energy damping (growing) rates i.e $-Q_p$ ($Q_p$).}}
    \label{fig:wave_diss_rate}
\end{figure*}
Upon evaluating dissipation rates of ion-scale waves by employing the routine described in Section \ref{sec:methodology} for all intervals from PSP Encounters 1 and 2 between 0.166 and 0.25 au that were analyzed in SKM23, we find that ICWs can dissipate onto protons at significant rates that could explain the anisotropic heating of protons which enhances at closer distances to the Sun \citep{Hellinger_2011, Hellinger_2013}. 
Figure \ref{fig:wave_diss_rate} shows the dissipation rates onto protons, $Q_p$ by cyclotron damping of ion-scale wave estimated in this work and Landau damping ($Q_{p,Landau}$, green) of turbulence estimated in SKM23. Here dots represent net damping rates and open circles represent net growth rates. Panels (a,b) show $Q_p$ via cyclotron damping of positively (red) and negatively (blue) Doppler-shifted ICWs and the corresponding moving averages/medians (solid lines). The mean $Q_p$ values via cyclotron damping are usually an order of magnitude higher than $Q_p$ values via Landau damping and are in good agreement with the inertial scale energy cascade rates estimated at 36 and 54 solar radii by \cite{Bandyopadhyay_2020} using PSP observations (blue stars). That the cyclotron damping rate is similar to the turbulent cascade rate is consistent with \cite{Bowen2024}, who show that ion-scale waves are correlated with signatures of sub-ion-scale turbulence and suggest that the waves play an active role in dissipating turbulence. Note that for $Q_p$ via cyclotron resonance the mean is an order of magnitude higher than the median while for $Q_p$ via Landau resonance the mean and median are similar. We infer that the proton cyclotron damping is intermittent and bursty, while the proton Landau damping is constant and continuous. We find that in \revise{12.74\%} of the intervals where waves are observed, there is a net ICW energy growth rate (open circles in panel (a)). This fraction is lower than the fraction of intervals (38\%) where the interval averaged values of $\frac{T_{\perp,p}}{T_{\parallel,p}}$ determine a growing wave at any wavevector. This discrepancy arises because ion-scale waves are observed over a wide effective $k_\parallel$ range, while the instabilities due to temperature anisotropy arise over a narrower $k_\parallel$ band (as seen in Figure \ref{fig:gamma_omega_spectrogram}). 
Integrating the growth and damping rates over the observed scales often results in a net damping rate.

We define a total possible heating rate due to cyclotron damping onto protons, $Q_{p,cyc}$, as the sum of the red and blue dots in panel (a). In intervals where both LH and RH polarized waves are observed, two distinct oppositely propagating ICWs can be positively and negatively Doppler shifted to independently produce both the observed polarizations, and there is no double counting in the evaluation of $Q_{p,cyc}$ by adding heating rates of both polarizations.

Moreover, we find that consistent with expectation, the possible dissipation of FMWs onto electrons at the observed scales is insignificant compared to $Q_{p,cyc}$ as well as $Q_{e,Landau}$, the turbulence dissipation rates onto electrons via Landau damping that were estimated in SKM23. 

It is important to note that in this work we have assumed the proton velocity distribution to be a bi-Maxwellian. However, the observed proton velocity distributions have beams. While accounting for the beams in the fits of VDFs would enhance the proton core temperature anisotropy estimates, \citep{Huang_2020, Halekas2020} the presence of beams can drive instabilities \citep{Montgomery_1975, Klein_2018_beam_instability, Mihailo_2021_instabilities_Helios} that are not considered in this work. We subsequently do not consider the interaction of proton beams with FMWs. However, FMW-ion interaction is relevant to the pitch angle diffusion of the ion beams \citep{Verniero_2022, Martinovic_2023_SAVIC}. Accounting for beams in this kind of analysis is a topic for future work.

\begin{figure}[hbt!]
    \centering
    \includegraphics[scale = 0.29]{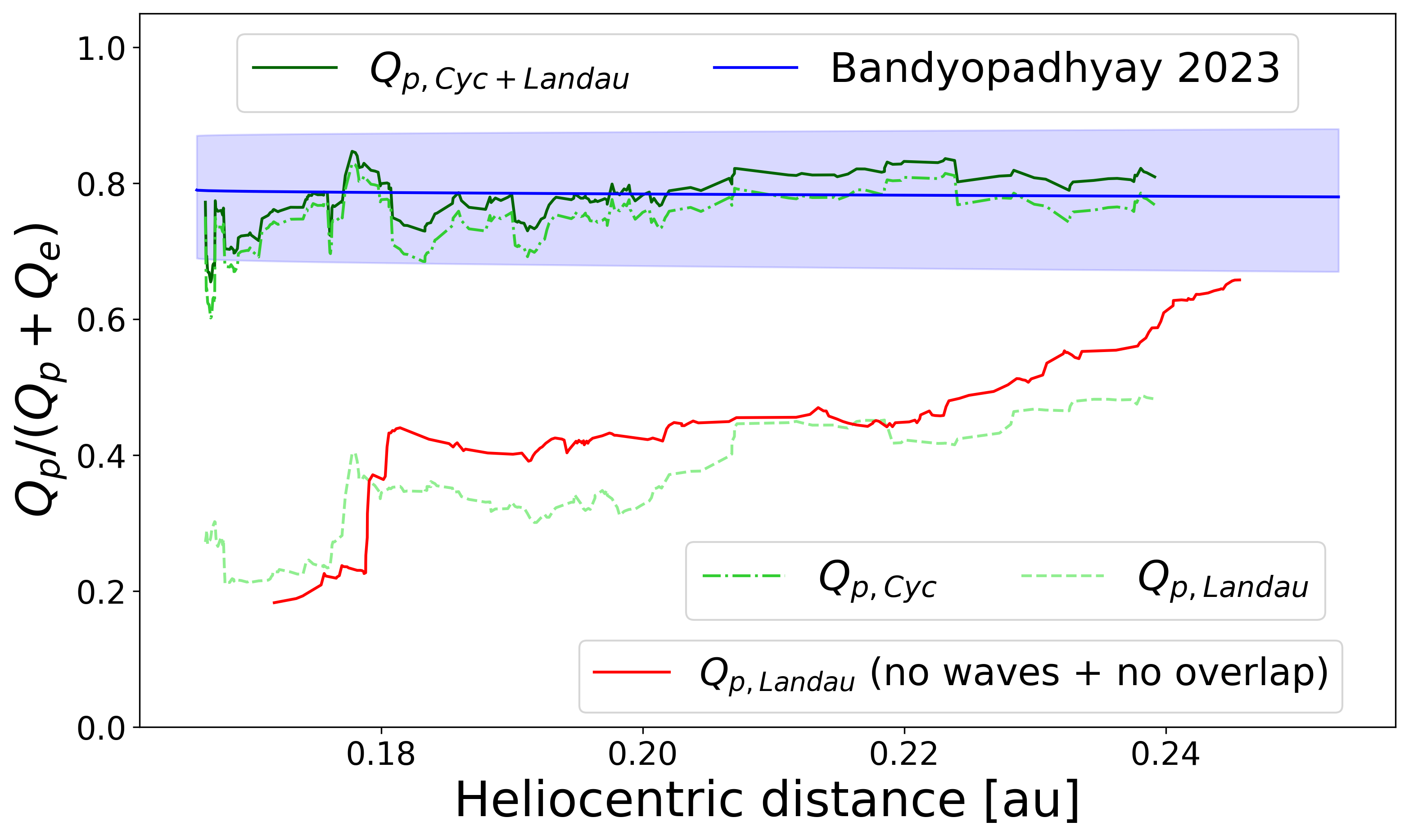}
    \caption{\revise{Predicted proton-to-total heating rates $Q_p/(Q_p + Q_{e})$ as a function of heliospheric distance evaluated by calculating the moving averages of $Q_p$ and $Q_e$. 
    We separately consider proton heating via the Landau resonance (light green dashed), the cyclotron resonance (medium green dashed-dotted) and both resonances (dark green solid) in intervals where waves are observed (green shaded in Table \ref{tab:Cyc_Ld_availability}). We also show $Q_p/(Q_p + Q_{e})$ evaluated from Landau damping alone (red) in intervals where waves are not observed (red shaded in Table \ref{tab:Cyc_Ld_availability}). The fit of $Q_p/(Q_p + Q_{e})$ along with error bars estimated by \cite{Bandy_2023} is shown (cyan) for comparison.}}
    \label{fig:Q_p_Q_total_ratio}
\end{figure}

\subsection{$Q_p/(Q_p + Q_e)$ consistent with the dominant proton heating that is observed near the Sun}

Because cyclotron damping onto electrons is negligible, we evaluate $Q_p/(Q_p + Q_e)$ by considering just $Q_{e,Landau}$ for electrons while we consider both $Q_{p,cyc}$ and $Q_{p,Landau}$ for protons. Moreover the estimations of $Q_{p/e,Landau}$ and $Q_{p,cyc}$ are only possible in intervals where the cascade model employed in SKM23 is accurate and waves are present, respectively. 
\revise{Additionally, the cascade model assumes a distribution of turbulent energy with large wavevector anisotropy perpendicular to \textbf{B}.
As the angle between $\mathbf{V}_{SW}$ and \textbf{B} decreases, the observed energy spectra is increasingly dominated by any structures parallel to \textbf{B} (e.g. Figure 2 in \cite{Horbury_2008}), and insensitive to the perpendicular turbulence.
For the comparison of heating rates in this work, we only consider Landau damping estimates in intervals where the perpendicular structure is resolved, $\theta_{BV_{SW}} > 30^\circ$, discarding approximately half of the intervals where the Landau Damping model was considered accurate in SKM23}.
Table \ref{tab:Cyc_Ld_availability} shows the percentages of the analyzed intervals where $Q_{p/e,Landau}$ and $Q_{p,cyc}$ estimations are available.

\begin{table}[hbt!]
\setlength{\tabcolsep}{8pt}
    \centering
    \begin{tabular}{m{0.8cm} m{2.75cm}| c c}
    \hline
    \multirow{2}{1cm}{} & & \multicolumn{2}{c}{Landau damping model}
    \\ [1 ex] 
    & & \multicolumn{1}{c}{accurate} & \multicolumn{1}{c}{inaccurate}  \\
    \hline 
    \multirow{2}{1.5cm}{Waves} &  present & \cellcolor{green!15} 15.3 &  55.31  \\ [1 ex]
    
    & absent/no overlap & \cellcolor{red!15} 12.79 & 5.88  \\
    \hline
    \end{tabular}
    \caption{\revise{Percentages of intervals where Landau and cyclotron damping rates are available. Landau damping rates are available when the cascade model employed in SKM23 is accurate. Cyclotron damping rates are available when waves are observed and the frequencies at which they are observed, $\omega_{\text{obs}}$ overlap with the Doppler shifted \texttt{PLUME} frequencies, $\omega_{sc}$.}}
    \label{tab:Cyc_Ld_availability}
\end{table}

We consider intervals where both  $Q_{p/e,Landau}$ and $Q_{p,cyc}$ available (i.e. \revise{15.3\%} of intervals, green shaded in Table \ref{tab:Cyc_Ld_availability}) and evaluate moving averages of $Q_p$ and $Q_e$ and then evaluate $Q_p/(Q_p + Q_{e})$ (shown in Figure \ref{fig:Q_p_Q_total_ratio}) corresponding to $Q_{p,cyc}$ (medium green dashed-dotted), $Q_{p,Landau}$ (light green dashed), and their sum (dark green solid) as functions of heliocentric distance. We infer that cyclotron damping greatly enhances the dissipation onto protons, producing much higher rates than those possible by Landau damping alone. By using PSP observations, \cite{Bandy_2023} have extended to near Sun distances the $Q_{p}/(Q_p + Q_e)$ estimation made by \cite{Cranmer_2009} using Ulysses and Helios observations. They estimate that protons receive $\sim$ 80\% of plasma heating (cyan in panel (a) of Figure \ref{fig:Q_p_Q_total_ratio}). 
Our estimates of $Q_p/(Q_p + Q_{e})$ (dark green) are in good agreement with theirs.

 In \revise{55.31\% of intervals, ion-scale waves are observed but the cascade model employed to estimate $Q_{e,Landau}$ is inaccurate. The mean value of $Q_{p,cyc}$ is the same in intervals where the Landau damping model is or is not accurate.} In 5.2\% of analyzed intervals, there is no overlap between $\omega_{\text{obs}}$ and $\omega_{sc}$ which occurs because of discarding heavily damped \texttt{PLUME} solutions.

\section{Discussion}
\label{sec:discussion}
We identify and analyze the abundantly observed ion-scale waves in Encounters 1 and 2 of PSP measurements and account for their dissipation via cyclotron damping. Along with our previous work, which accounts for the dissipation of turbulence via Landau damping, our findings help to better quantify the mechanisms that heat the young solar wind as it expands.

 We find that ion-scale waves are observed over 20.37$\%$ of PSP Encounters 1 and 2 observational time of which the LH polarized (in spacecraft frame) waves are dominant constituting 85.86$\%$ of the time during which waves are observed. The plasma-frame frequencies corresponding to the observed waves indicate that the waves are locally generated ICWs. We further infer that these ICWs are transient with decay times comparable to the turbulent non-linear interaction times at proton scales. ICWs among the observed waves could thus mediate the local transport of free energy associated with temperature anisotropy. We find pronounced dissipation of ICWs onto protons that could account for the enhanced anisotropic heating of protons near the Sun and the estimated rates of dissipation are in good agreement with empirical estimates of turbulent energy cascade rates by \cite{Bandyopadhyay_2020}. We further infer dominant proton heating and the ratio of proton-to-total heating estimated in this work is in good agreement with the estimate by \cite{Bandy_2023}.

Intermittent ion-scale instabilities are a likely mechanism to generate ICWs. We find a non-linear correlation of wave occurrence ($T_{wave}$) and $\frac{T_{\perp,p}}{T_{\parallel,p}}$ (Spearman correlation coeffecient = 0.37). As the waves are short-lived, these 15-minute averaged $\frac{T_{\perp,p}}{T_{\parallel,p}}$ values may not well represent the anisotropy of the shorter intervals where the waves would be generated, preventing a conclusive inference from this correlation.

Spectral steepening is often observed in Encounters 1 and 2 (\cite{Bowen_spectral_steepening, Shankarappa_2023}) at ion gyroscales. We find a strong correlation of $T_{wave}$ with the steepening of the energy spectrum. Bowen et al. 2024 (in review) find that this steepening is strongly associated with LH ion-scale waves. Furthermore, \cite{Squire_2022} have argued that the imbalanced turbulent flux creates a ``helicity barrier'' at ion gyroscales, inhibiting further cascade to smaller $k_\perp$, leading to an inverse cascade for $k_\perp > 1$ channeling the turbulent energy to oblique ICWs that dissipate onto protons leading to a steeper energy spectrum near the proton gyroscale. Although we do not find a notable correlation between $T_{wave}$ and cross-helicity extracted from our 15 minute intervals, this is an area of future research.

A significant caveat in this work is the assumption that bi-Maxwellians well represent the proton and electron VDFs. The presence of proton beams can drive instabilities which is unaccounted for in this work. 
Moreover, the structure of VDFs can differ significantly from Maxwellian \citep{Marsch_1982_proton_VDFs,Bowen_2022_Cyclotron_dissipation, Walters_2023}; consideration of more accurate solar wind particle VDFs and their impact on the dissipation and emission of waves is a topic for future work. Additionally, we have not taken into account the possibility of proton velocity distributions reaching marginal stability and thus being unable to further exchange energy with the ion-scale waves \citep{Isenberg_2000}.

\acknowledgments
KGK and NS were supported by NASA Grants 80NSSC19K0912 and 80NSSC20K0521. MMM and KGK were supported by NASA grant 80NSSC22K1011. TAB acknowledges support from NASA via grant 80NSSC24K0272.

Parker Solar Probe was designed, built, and is now operated by the Johns Hopkins Applied Physics Laboratory as part of NASA's Living with a Star (LWS) program (contract NNN06AA01C). Support from the LWS management and technical team has played a critical role in the success of the Parker Solar Probe mission. 

\appendix

\section{Identifying $\sigma(s,t)$ of Ion-scale Waves by Removing non-wave signatures} \label{App:wave_identification}

We identify ion-scale waves from turbulence due to the persistent high values of spacecraft frame polarization, $\sigma$ of the former as compared to the random short-lived low $\sigma$ values due to the latter. Turbulent fluctuations can often have random high values of $\sigma$ and can lead to their erroneous identification as waves when $\sigma$ is used as a criterion. In order to remove these short-lived random contributions to the $\sigma$ and retain only the persistent high-polarized bands, we perform a moving window time averaging on $\sigma(s,t)$ with a window width of $20 \times \frac{1}{\text{frequency}}$ at all $f_{\text{log\_bin}}$ frequencies ranging from 0.005 Hz to 10 Hz. 
The averaging factor of 20 is empirically determined by identifying the minimum window width required to reduce magnitudes of averaged $|\sigma(s,t)|$ to less than 0.7 at all times and frequencies in intervals where ion-scale waves are not observed. After time averaging, we identify domains in frequency-time space where $|\sigma(s,t)| \ge 0.7$ (the choice of the threshold value of 0.7 is discussed in Appendix B.4. of SKM23), $(f-t)_{coherent}$, which we attribute to the presence of coherent waves (depicted in panel (b) of Figure \ref{fig:wave_diss_methodology}). The waves are often observed over a wide range of frequencies. 

Furthermore, in SKM23, each analyzed 15-minute interval was extended by considering 7.5-minute peripheral intervals on either end to limit finite-length errors (see Appendix B.3.2 of SKM23). We employed a method to remove reaction wheel noise (see Appendix B.1 of SKM23) that was based on the relative magnitudes of energies in the central 15-minute interval and the peripheral 7.5-minute intervals. The noise-removal method chooses between the following two routines: removing the noise in the whole extended interval, or removing the noise in central and peripheral intervals separately and then joining them. This method is not effective in a handful of intervals where the energies of central and peripheral intervals are similar. In many intervals, a narrow frequency band, $f_{noise}$, with high $\sigma$ values corresponding to the remnant noise, persists across a region in $(f-t)_{coherent}$ despite time averaging. The frequencies of $f_{noise}$ are usually higher than and thus do not overlap with, the ion-scale wave frequencies and have lower amplitude values of their $\sigma$ compared to the latter. We identify $f_{noise}$ as the frequencies in $(f-t)_{coherent}$ with the time-averaged spacecraft frame polarization value, $\left<|\sigma(f,t)|\right>_t$ is less than an empirical threshold of 0.0078. We can then remove $f_{noise}$ and retain noise-removed $(f-t)_{coherent}$. The noise-frequency filtering additionally removes the edge frequencies in $(f-t)_{coherent}$ that are far from the core region of ion-scale wave frequencies. However, we have found that the energies in these edge frequencies are negligible and don't affect our analysis.

We consider the wavelet energy spectrum (evaluated as in SKM23, see Appendix B.3) corresponding to RH and LH polarization at $(f-t)_{coherent}$ and sum these domains over time to evaluate $|\tilde{B}(f)|^{2^{LH/RH}}_{wave\_psp}$ (shown in panel (c) of Figure \ref{fig:wave_diss_methodology}). 
We consider the frequency range of $(f-t)_{coherent}$ domain as the frequencies where waves are observed in the spacecraft frame, $f^{(RH/LH)}_{\text{obs}}$, and evaluate the corresponding angular frequencies, $\omega^{(RH/LH)}_{\text{obs}}$. Note that choosing a $\sigma$ threshold value of > 0.7 in Section \ref{subsec_sigma_smooth} causes an insignificant difference in the evaluation of $f_{obs}^{LH/RH}$.

\section{Identification of modes in \texttt{PLUME}} \label{app:mode_identify}
We accurately identify the four modes and distinguish the Alfv\'en ICW from FMW by applying the following criteria in succession,
\begin{enumerate}
\item For a plasma with no drifting components, the entropy mode is non-propagating at large scales, i.e. $\omega = 0$ for $k_\parallel \rho_p \ll 1$, allowing immediate identification of the entropy mode.
\item For the three propagating modes, the variation in the parallel group velocity, $v_{g,\parallel}= \frac{\partial \omega}{ \partial k_\parallel}$ offers a measure to distinguish between solutions. 
The values of $v_{g,\parallel}$ increase, decrease, and remain roughly constant for the fast, Alfv\'en, and slow waves, respectively with increasing $k_\parallel$; see Equations 8, 14, 16, and 18 in \cite{Howes_2014_Taylor_Hypothesis} for analytic approximations of $\omega(k)$ for Alfv\'en and FMWs at inertial and parallel kinetic scales. 
At small parallel scales, the ICWs damp heavily onto ions, resulting in a decrease in $v_{g,\parallel}$, while the FMWs don't damp onto ions; their frequency scales as $\propto k k_\parallel$ resulting in an increase in $v_{g,\parallel}$. 
We calculate $ \Delta v_{g,\parallel}$, the change in the parallel group velocity between $k_\parallel \rho_p = 0.003$ and $k_\parallel \rho_p = 1$ for the three modes. 
The mode with the maximum positive $ \Delta v_{g,\parallel}$ value is identified as the FMW, and the one with maximum negative $ \Delta v_{g,\parallel}$ value is identified as the Alfv\'en ICW mode.
\item To verify the above identification, we utilize two additional wave metrics, magnetic helicity (see Equation \ref{eq:magnetic_helicicty} in the following section) which quantifies the helicity of the wave mode, and the ratio of magnitudes of parallel electric field to total electric field, $\frac{E_z}{E}$. 
The forward propagating FMW and Alfv\'en waves have a constant $\sigma_{m}$ value of +1 and -1, respectively for $k_\parallel \rho_p > 0.01$, while the slow wave has $\sigma_{m}$ values varying between +1 and -1. 
Only the slow mode has a $\frac{E_z}{E}$ value of $\approx 1$ at larger parallel scales, while both Alfv\'en and FMWs have a value of $\approx 0$ for the considered range of wavevectors. 
We verify that the identified FMW and Alfv\'en waves have expected magnitudes of both the wave metrics at two values of parallel scales, $k_\parallel \rho_p = 0.1$ and $ k_\parallel \rho_p = 2$, further ensuring that there is no numerical mode interchange near $k_\parallel \sim \rho_p$.
\end{enumerate}

\section{Quantifying circular Polarization of waves} \label{app:polarization}
ICWs and FMWs are circularly polarized at scales of order $\rho_p$ and in their plasma dispersion relations, $\left(\frac{k}{\omega} \right)^2_{\text{ICW},\text{FMW}}$ is a function of $\left( \frac{1}{\omega - \mathbf{k}\cdot \mathbf{v} - \Omega_{p,e} }\right) $, where $\Omega_{p,e} = \frac{eB}{m_{p,e}}$ is the proton (electron) cyclotron frequency. Hence, The ICWs (FMWs) propagating both along and opposite to \textbf{B} resonate with protons (electrons) in the plasma rest frame when beams are not considered. Note that the handedness of light has conventionally been defined with respect to the observer's point of view (i.e. source or receiver). A rotating magnetic field of a wave can appear clockwise (anti-clockwise) when viewed from the source (receiver) point of view. However, in this work (following \cite{Bowen_ICW}), we avoid this ambiguity by considering proton (electron)-resonant polarization to be left (right)-handed irrespective of the observer's point of view. Hence, ICWs (FMWs) propagating both along and opposite to \textbf{B} are left (right)-handed i.e. LH (RH). 

Two interrelated metrics quantify circular polarization in waves: magnetic helicity ($\sigma_m$) and polarization/reduced helicity spectrum ($\sigma$).
\begin{itemize}
    \item \textbf{Magnetic helicity}, $\sigma_m = \int d^3 x \mathbf{A}\cdot \mathbf{B}$ is an invariant of ideal MHD equations, where \textbf{A} is the vector potential of \textbf{B}. $\sigma_m$ is a good measure of circular polarization when \textbf{B} is wavelike. In the wavevector space, normalized  magnetic helicity for $k \eqsim k_z$ is written as, \citep{Matthaeus_1982_rugged_invariants,Howes_2010_helicity}, 
    \begin{equation} \label{eq:magnetic_helicicty}
    \sigma_{m} (k) = \frac{k}{|B (\mathbf{k})|^2} i \left(\frac{B_x(k) B^{*}_{y}(k) -  B^{*}_{x}(k)   B_y(k)}{k_z}\right) = \frac{-2\text{Im}\left[B_x(k) B_y^*(k)\right]}{|B_x(k)|^2 + |B_y(k)|^2}..
\end{equation}
Here $B_{x,y}(k)$ are spatial Fourier transforms of fluctuations of the magnetic field, $\delta B$. 

\item \textbf{Polarization}, $\sigma$ is the ratio of third-to-zeroth Stokes parameters and is written as \citep{Bowen_ICW},
\begin{equation} \label{eq:coherence}
 \sigma (f) = \dfrac{-2  \text{Im}\big[B_x(f)B_y^*(f)\big]}{|B_x(f)|^2 + |B_y(f)|^2} \ \ \ \ \ \text{or} \ \ \ \ \ \sigma (s,t) = \dfrac{-2  \text{Im}\big[W_x(s,t)W_y^*(s,t)\big]}{|W_x(s,t)|^2 + |W_y(s,t)|^2}.
\end{equation}
Here $B(f)/W_{x,y}(s,t)$ are temporal Fourier/wavelet transforms of $\delta B$. Polarization, effectively is the reduced magnetic helicity spectrum along a single direction \citep{Matthaeus_1982_rugged_invariants}. We can measure only polarization and not magnetic helicity with single spacecraft measurements of magnetic field observations.

\end{itemize}

\begin{figure}[h]
    \centering
    \plotone{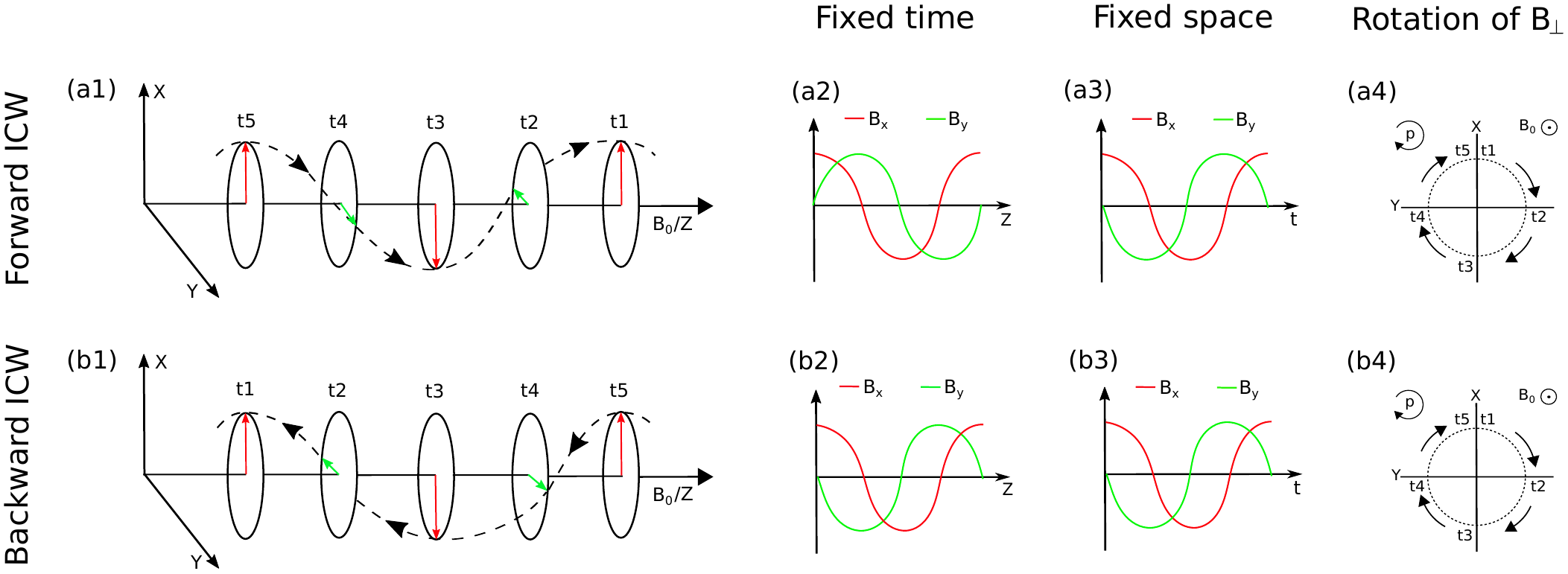}
    \caption{Spatial and temporal behavior of magnetic field components in the plane perpendicular to the direction of propagation of circularly polarized counter-propagating ICWs in the plasma rest frame.}
    \label{fig:Helicity_vs_sigma}
\end{figure}

In the plasma frame, magnetic helicity quantifies the rotation of \textbf{B} with respect to space at a fixed time while polarization quantifies the rotation of \textbf{B} with respect to time at a fixed spatial point \citep{Narita_2009}. Magnetic helicity is defined with respect to the wavevector, \textbf{k}, while polarization is not. Hence, in the plasma rest frame, magnetic helicity (polarization) has the opposite (same) sign for counter-propagating waves with the same intrinsic polarization. This difference is depicted in Figure \ref{fig:Helicity_vs_sigma} which is a variation of Figure 1 from \cite{Narita_2009} (where a RH wave is depicted, as opposed to the LH wave in our example). Panels (a1) and (b1) show forward and backward propagating example ICWs in the plasma rest frame where black arrows show the direction of wave propagation. Here XYZ is a right coordinate system where the mean magnetic field, $\mathbf{B_o}$ is in the \textbf{Z} direction. The position of $\mathbf{ B_\perp} = \mathbf{B_x} + \mathbf{B_y}$ in the plane perpendicular to $\mathbf{B_o}$ is shown as colored arrows at 5 consecutive points along \textbf{Z}. When $\mathbf{ B_\perp}$ points towards \textbf{X} or \textbf{Y} directions, the arrows are colored red or green, respectively.  

When the $\mathbf{ B_\perp}$ of waves is observed with respect to time at a fixed spatial point, the observer sees it in the order t1-t5 for both cases and the rotation of $\mathbf{B_\perp}$ (as shown in panels (a4,b4)) is observed to be proton-resonant (LH) for both the waves. The corresponding oscillations of $B_x$ and $B_y$ with respect to time (shown in panels (a3,b3)) are identical which is consistent with the sign of polarization for both the forward and backward ICWs being the same. On the other hand, panels (a2,b2) show the oscillations of $B_x$ and $B_y$ with respect to space (z) at a fixed time, and the variation of $B_y$ relative to $B_x$ is inverted for the forward and backward waves which is consistent with the sign of magnetic helicity being reversed.

 The expressions for magnetic helicity (polarization) (Equations \ref{eq:magnetic_helicicty} and \ref{eq:coherence}) evaluate the phase difference between $B_x$ and $B_y$ corresponding to their spatial (temporal) variations. This can be shown by expressing the oscillating magnetic field of forward/backward propagating electromagnetic wave as complex numbers $B_{x,y} \propto e^{i(kz \mp \omega t + \theta_{x,y})}$, where $\theta_{x,y}$ are phases of $B_{x,y}$. In length or time space, 
$$\dfrac{-2  \text{Im}\big[B_xB_y^*\big]}{|B_x|^2 + |B_y|^2} = - \text{Im} \left( e^{i( kz \mp \omega t + \theta_x)} e^{-i(kz \mp \omega t + \theta_y)} \right)  =  \sin (\theta_y - \theta_x) .$$
 This phase difference between $B_x$ and $B_y$ is conserved in the Fourier wavevector (frequency) domain. The values of magnetic helicity and polarization can be calculated for the example forward/backward ICWs by using expressions for $B_x$ and $B_y$ as $B_{x} \propto \cos{(kz \mp \omega t )}$, and $B_{y} \propto \cos{(kz \mp \omega t \mp \pi/2)}$, respectively. Table \ref{tab:helicity_coherence} summarizes the values of $\sigma_m$ and $\sigma$ for the forward and backward ICW and FMWs in the plasma frame.  Polarization observed by spacecraft is Doppler-shifted and can change signs due to the relative motion of the spacecraft with respect to the solar wind, as explained in Section \ref{sec:Doppler_shifts}.

\begin{table}
\setlength{\tabcolsep}{8pt}
    \centering
    \begin{tabular}{c c| c c c}
    \hline
    & & magnetic helicity & polarization\\[1 ex] 
    \hline
    \multirow{2}{0.75cm}{ICW} & Forward & -1 & 1\\
    & Backward & 1 & 1\\
    \hline 
    \multirow{2}{0.75cm}{FMW} & Forward & 1 & -1\\
    & Backward & -1 & -1\\
    \hline
    \end{tabular}
    \caption{Magnetic helicity ($\sigma_m$) and polarization($\sigma$) values for forward and backward propagating (with respect to \textbf{B}) ICW and FMWs in plasma rest frame.}
    \label{tab:helicity_coherence}
\end{table}

\bibliographystyle{apj}
\bibliography{References.bib}

\begin{thebibliography}{}

\bibitem[\protect\citeauthoryear{{Bale} et~al.}{{Bale} et~al.}{2016}]{Bale2016}
{Bale}, S.~D., et~al. 2016, Space Sci.~Rev., 204, 49

\bibitem[\protect\citeauthoryear{Bale et~al.}{Bale et~al.}{2009}]{Bale_2009_PRL_T_anisotropy_instabilities}
Bale, S.~D., Kasper, J.~C., Howes, G.~G., Quataert, E., Salem, C.,  \& Sundkvist, D. 2009, Phys. Rev. Lett., 103, 211101

\bibitem[\protect\citeauthoryear{Bandyopadhyay et~al.}{Bandyopadhyay et~al.}{2020}]{Bandyopadhyay_2020}
Bandyopadhyay, R., et~al. 2020, The Astrophysical Journal Supplement Series, 246, 48

\bibitem[\protect\citeauthoryear{{Bandyopadhyay} et~al.}{{Bandyopadhyay} et~al.}{2023}]{Bandy_2023}
{Bandyopadhyay}, R., et~al. 2023, \apjl, 955, L28

\bibitem[\protect\citeauthoryear{Barnes}{Barnes}{1966}]{Barnes_1966}
Barnes, A. 1966, The Physics of Fluids, 9, 1483

\bibitem[\protect\citeauthoryear{Batchelor}{Batchelor}{1953}]{Batchelor1953}
Batchelor, G.~K. 1953, Quarterly Journal of the Royal Meteorological Society, 79, 457

\bibitem[\protect\citeauthoryear{{Bowen} et~al.}{{Bowen} et~al.}{2020}]{Bowen_Outward_ICW_2020}
{Bowen}, T.~A., et~al. 2020, \apj, 899, 74

\bibitem[\protect\citeauthoryear{Bowen et~al.}{Bowen et~al.}{2024}]{Bowen2024}
Bowen, T.~A., et~al. 2024, Nature Astronomy, 8, 482

\bibitem[\protect\citeauthoryear{Bowen et~al.}{Bowen et~al.}{2022}]{Bowen_2022_Cyclotron_dissipation}
Bowen, T.~A., et~al. 2022, Phys. Rev. Lett., 129, 165101

\bibitem[\protect\citeauthoryear{Bowen et~al.}{Bowen et~al.}{2020a}]{Bowen_spectral_steepening}
Bowen, T.~A., et~al. 2020a, Phys. Rev. Lett., 125, 025102

\bibitem[\protect\citeauthoryear{Bowen et~al.}{Bowen et~al.}{2020b}]{Bowen_ICW}
Bowen, T.~A., et~al. 2020b, The Astrophysical Journal Supplement Series, 246, 66

\bibitem[\protect\citeauthoryear{{Case} et~al.}{{Case} et~al.}{2020}]{Case_2020}
{Case}, A.~W., et~al. 2020, Astrophys.~J.~Supp., 246, 43

\bibitem[\protect\citeauthoryear{Chandran et~al.}{Chandran et~al.}{2010}]{Chandran_2010}
Chandran, B. D.~G., Li, B., Rogers, B.~N., Quataert, E.,  \& Germaschewski, K. 2010, The Astrophysical Journal, 720, 503

\bibitem[\protect\citeauthoryear{Chen, Klein, \& Howes}{Chen et~al.}{2019}]{Chen_2019_LD}
Chen, C., Klein, K.,  \& Howes, G. 2019, Nature Communications, 10

\bibitem[\protect\citeauthoryear{Chen, Lin, \& White}{Chen et~al.}{2001}]{Chen_2001_SH}
Chen, L., Lin, Z.,  \& White, R. 2001, Physics of Plasmas, 8, 4713

\bibitem[\protect\citeauthoryear{{Cranmer} et~al.}{{Cranmer} et~al.}{2009}]{Cranmer_2009}
{Cranmer}, S.~R., {Matthaeus}, W.~H., {Breech}, B.~A.,  \& {Kasper}, J.~C. 2009, \apj, 702, 1604

\bibitem[\protect\citeauthoryear{Dmitruk, Matthaeus, \& Seenu}{Dmitruk et~al.}{2004}]{Dmitruk_2004}
Dmitruk, P., Matthaeus, W.~H.,  \& Seenu, N. 2004, The Astrophysical Journal, 617, 667

\bibitem[\protect\citeauthoryear{{Fox} et~al.}{{Fox} et~al.}{2015}]{Fox:2015}
{Fox}, N.~J., et~al. 2015, Space Sci.~Rev.

\bibitem[\protect\citeauthoryear{Gary}{Gary}{1993}]{Gary_SP_1993}
Gary, S.~P. 1993, Theory of Space Plasma Microinstabilities, Cambridge Atmospheric and Space Science Series (Cambridge University Press)

\bibitem[\protect\citeauthoryear{Goldreich \& Sridhar}{Goldreich \& Sridhar}{1995}]{GS1995}
Goldreich, P.,  \& Sridhar, S. 1995, Astrophys.~J., 438, 763

\bibitem[\protect\citeauthoryear{{Halekas} et~al.}{{Halekas} et~al.}{2020}]{Halekas2020}
{Halekas}, J.~S., et~al. 2020, Astrophys.~J.~Supp., 246, 22

\bibitem[\protect\citeauthoryear{He et~al.}{He et~al.}{2015}]{He_2015}
He, J., Wang, L., Tu, C., Marsch, E.,  \& Zong, Q. 2015, The Astrophysical Journal Letters, 800, L31

\bibitem[\protect\citeauthoryear{Hellinger et~al.}{Hellinger et~al.}{2011}]{Hellinger_2011}
Hellinger, P., Matteini, L., {\v{S}}tver{\'a}k, {\v{S}}., Tr{\'a}vn{\'i}{\v{c}}ek, P.~M.,  \& Marsch, E. 2011, Journal of Geophysical Research: Space Physics, 116

\bibitem[\protect\citeauthoryear{Hellinger et~al.}{Hellinger et~al.}{2006}]{Hellinger_2006}
Hellinger, P., Tr{\'a}vn{\'i}{\v{c}}ek, P., Kasper, J.~C.,  \& Lazarus, A.~J. 2006, Geophysical Research Letters, 33

\bibitem[\protect\citeauthoryear{Hellinger et~al.}{Hellinger et~al.}{2013}]{Hellinger_2013}
Hellinger, P., Tr{\'a}vn{\'i}{\v{c}}ek, P.~M., {\v{S}}tver{\'a}k, {\v{S}}., Matteini, L.,  \& Velli, M. 2013, Journal of Geophysical Research: Space Physics, 118, 1351

\bibitem[\protect\citeauthoryear{Hollweg \& Isenberg}{Hollweg \& Isenberg}{2002}]{Hollweg_Isenberg_ICW_diss}
Hollweg, J.~V.,  \& Isenberg, P.~A. 2002, Journal of Geophysical Research: Space Physics, 107, SSH 12

\bibitem[\protect\citeauthoryear{Horbury, Forman, \& Oughton}{Horbury et~al.}{2008}]{Horbury_2008}
Horbury, T.~S., Forman, M.,  \& Oughton, S. 2008, Phys. Rev. Lett., 101, 175005

\bibitem[\protect\citeauthoryear{Howes et~al.}{Howes et~al.}{2008}]{Howes2008}
Howes, G.~G., Cowley, S.~C., Dorland, W., Hammett, G.~W., Quataert, E.,  \& Schekochihin, A.~A. 2008, Journal of Geophysical Research: Space Physics, 113

\bibitem[\protect\citeauthoryear{{Howes}, {Klein}, \& {TenBarge}}{{Howes} et~al.}{2014}]{Howes_2014_Taylor_Hypothesis}
{Howes}, G.~G., {Klein}, K.~G.,  \& {TenBarge}, J.~M. 2014, \apj, 789, 106

\bibitem[\protect\citeauthoryear{{Howes} \& {Quataert}}{{Howes} \& {Quataert}}{2010}]{Howes_2010_helicity}
{Howes}, G.~G.,  \& {Quataert}, E. 2010, The Astrophysical Journal Letters, 709, L49

\bibitem[\protect\citeauthoryear{Huang et~al.}{Huang et~al.}{2020}]{Huang_2020}
Huang, J., et~al. 2020, The Astrophysical Journal Supplement Series, 246, 70

\bibitem[\protect\citeauthoryear{{Isenberg}, {Lee}, \& {Hollweg}}{{Isenberg} et~al.}{2000}]{Isenberg_2000}
{Isenberg}, P.~A., {Lee}, M.~A.,  \& {Hollweg}, J.~V. 2000, \solphys, 193, 247

\bibitem[\protect\citeauthoryear{Jian et~al.}{Jian et~al.}{2009}]{Jian_2009}
Jian, L.~K., Russell, C.~T., Luhmann, J.~G., Strangeway, R.~J., Leisner, J.~S.,  \& Galvin, A.~B. 2009, The Astrophysical Journal, 701, L105

\bibitem[\protect\citeauthoryear{{Kasper} et~al.}{{Kasper} et~al.}{2016}]{Kasper2016}
{Kasper}, J.~C., et~al. 2016, Space Sci.~Rev., 204, 131

\bibitem[\protect\citeauthoryear{Kasper et~al.}{Kasper et~al.}{2013}]{Kasper_2013_ICW_damping}
Kasper, J.~C., Maruca, B.~A., Stevens, M.~L.,  \& Zaslavsky, A. 2013, Phys. Rev. Lett., 110, 091102

\bibitem[\protect\citeauthoryear{Klein}{Klein}{2013}]{Klein_PhD_thesis}
Klein, K.~G. 2013, Ph.D. thesis, University of Iowa

\bibitem[\protect\citeauthoryear{{Klein} et~al.}{{Klein} et~al.}{2018}]{Klein_2018_beam_instability}
{Klein}, K.~G., {Alterman}, B.~L., {Stevens}, M.~L., {Vech}, D.,  \& {Kasper}, J.~C. 2018, \prl, 120, 205102

\bibitem[\protect\citeauthoryear{Klein \& Howes}{Klein \& Howes}{2015}]{Klein_Howes_2015_PLUME}
Klein, K.~G.,  \& Howes, G.~G. 2015, Physics of Plasmas, 22, 032903

\bibitem[\protect\citeauthoryear{{Klein} et~al.}{{Klein} et~al.}{2020}]{Klein_2020_JPP_Ld_Cyc_FPC}
{Klein}, K.~G., {Howes}, G.~G., {TenBarge}, J.~M.,  \& {Valentini}, F. 2020, Journal of Plasma Physics, 86, 905860402

\bibitem[\protect\citeauthoryear{{Leamon} et~al.}{{Leamon} et~al.}{1998}]{Leamon_1998_ICW}
{Leamon}, R.~J., {Matthaeus}, W.~H., {Smith}, C.~W.,  \& {Wong}, H.~K. 1998, Astrophys.~J.~Lett., 507, L181

\bibitem[\protect\citeauthoryear{Leamon et~al.}{Leamon et~al.}{1999}]{Leamon_1999}
Leamon, R.~J., Smith, C.~W., Ness, N.~F.,  \& Wong, H.~K. 1999, Journal of Geophysical Research: Space Physics, 104, 22331

\bibitem[\protect\citeauthoryear{Liu et~al.}{Liu et~al.}{2023}]{Liu_2023}
Liu, W., Zhao, J., Wang, T., Dong, X., Kasper, J.~C., Bale, S.~D., Shi, C.,  \& Wu, D. 2023, The Astrophysical Journal, 951, 69

\bibitem[\protect\citeauthoryear{Marsch et~al.}{Marsch et~al.}{1982}]{Marsch_1982_proton_VDFs}
Marsch, E., M{\"u}hlh{\"a}user, K.-H., Schwenn, R., Rosenbauer, H., Pilipp, W.,  \& Neubauer, F.~M. 1982, Journal of Geophysical Research: Space Physics, 87, 52

\bibitem[\protect\citeauthoryear{Martinovi{\'c} \& Klein}{Martinovi{\'c} \& Klein}{2023}]{Martinovic_2023_SAVIC}
Martinovi{\'c}, M.~M.,  \& Klein, K.~G. 2023, The Astrophysical Journal, 952, 14

\bibitem[\protect\citeauthoryear{{Martinovi{\'c}} et~al.}{{Martinovi{\'c}} et~al.}{2021}]{Mihailo_2021_instabilities_Helios}
{Martinovi{\'c}}, M.~M., {Klein}, K.~G., {{\v{D}}urovcov{\'a}}, T.,  \& {Alterman}, B.~L. 2021, \apj, 923, 116

\bibitem[\protect\citeauthoryear{Matthaeus \& Goldstein}{Matthaeus \& Goldstein}{1982}]{Matthaeus_1982_rugged_invariants}
Matthaeus, W.~H.,  \& Goldstein, M.~L. 1982, Journal of Geophysical Research: Space Physics, 87, 6011

\bibitem[\protect\citeauthoryear{{Matthaeus} \& {Velli}}{{Matthaeus} \& {Velli}}{2011}]{Matthaeus_Velli_2011}
{Matthaeus}, W.~H.,  \& {Velli}, M. 2011, \ssr, 160, 145

\bibitem[\protect\citeauthoryear{Matthaeus et~al.}{Matthaeus et~al.}{1999}]{Matthaeus_1999}
Matthaeus, W.~H., Zank, G.~P., Oughton, S., Mullan, D.~J.,  \& Dmitruk, P. 1999, The Astrophysical Journal, 523, L93

\bibitem[\protect\citeauthoryear{{McManus} et~al.}{{McManus} et~al.}{2024}]{Mcmanus_2024}
{McManus}, M.~D., et~al. 2024, \apj, 961, 142

\bibitem[\protect\citeauthoryear{Montgomery et~al.}{Montgomery et~al.}{1975}]{Montgomery_1975}
Montgomery, M.~D., Gary, S.~P., Forslund, D.~W.,  \& Feldman, W.~C. 1975, Phys. Rev. Lett., 35, 667

\bibitem[\protect\citeauthoryear{{Mozer} et~al.}{{Mozer} et~al.}{2023}]{Mozer_2023}
{Mozer}, F.~S., {Agapitov}, O.~V., {Kasper}, J.~C., {Livi}, R., {Romeo}, O.,  \& {Vasko}, I.~Y. 2023, \aap, 673, L3

\bibitem[\protect\citeauthoryear{Narita, Kleindienst, \& Glassmeier}{Narita et~al.}{2009}]{Narita_2009}
Narita, Y., Kleindienst, G.,  \& Glassmeier, K.-H. 2009, Annales Geophysicae, 27, 3967

\bibitem[\protect\citeauthoryear{{Quataert}}{{Quataert}}{1998}]{Quataert:1998}
{Quataert}, E. 1998, \apj, 500, 978

\bibitem[\protect\citeauthoryear{{Qudsi} et~al.}{{Qudsi} et~al.}{2020}]{Qudsi_2020}
{Qudsi}, R.~A., et~al. 2020, Astrophys.~J., 895, 83

\bibitem[\protect\citeauthoryear{Shankarappa, Klein, \& Martinović}{Shankarappa et~al.}{2023}]{Shankarappa_2023}
Shankarappa, N., Klein, K.~G.,  \& Martinović, M.~M. 2023, The Astrophysical Journal, 946, 85

\bibitem[\protect\citeauthoryear{{Sonnerup} \& {Cahill}}{{Sonnerup} \& {Cahill}}{1967}]{Sonnerup_1967_MVA}
{Sonnerup}, B.~U.~O.,  \& {Cahill}, J., L.~J. 1967, \jgr, 72, 171

\bibitem[\protect\citeauthoryear{{Squire} et~al.}{{Squire} et~al.}{2022}]{Squire_2022}
{Squire}, J., {Meyrand}, R., {Kunz}, M.~W., {Arzamasskiy}, L., {Schekochihin}, A.~A.,  \& {Quataert}, E. 2022, Nature Astronomy, 6, 715

\bibitem[\protect\citeauthoryear{Torrence \& Compo}{Torrence \& Compo}{1998}]{Terrence_compo}
Torrence, C.,  \& Compo, G.~P. 1998, Bulletin of the American Meteorological Society, 79, 61

\bibitem[\protect\citeauthoryear{Verniero et~al.}{Verniero et~al.}{2022}]{Verniero_2022}
Verniero, J.~L., et~al. 2022, The Astrophysical Journal, 924, 112

\bibitem[\protect\citeauthoryear{Verniero et~al.}{Verniero et~al.}{2020}]{Verniero_2020}
Verniero, J.~L., et~al. 2020, The Astrophysical Journal Supplement Series, 248, 5

\bibitem[\protect\citeauthoryear{Verscharen, Klein, \& Maruca}{Verscharen et~al.}{2019}]{verscharen2019}
Verscharen, D., Klein, K.~G.,  \& Maruca, B.~A. 2019, Living Reviews in Solar Physics, 16, 1

\bibitem[\protect\citeauthoryear{Walters et~al.}{Walters et~al.}{2023}]{Walters_2023}
Walters, J., Klein, K.~G., Lichko, E., Stevens, M.~L., Verscharen, D.,  \& Chandran, B. D.~G. 2023, The Astrophysical Journal, 955, 97

\bibitem[\protect\citeauthoryear{Whittlesey et~al.}{Whittlesey et~al.}{2020}]{Whittlesey_2020}
Whittlesey, P.~L., et~al. 2020, The Astrophysical Journal Supplement Series, 246, 74

\bibitem[\protect\citeauthoryear{{Woodham} et~al.}{{Woodham} et~al.}{2021}]{Woodham_2021}
{Woodham}, L.~D., {Wicks}, R.~T., {Verscharen}, D., {TenBarge}, J.~M.,  \& {Howes}, G.~G. 2021, \apj, 912, 101

\bibitem[\protect\citeauthoryear{{Zaslavsky}}{{Zaslavsky}}{2023}]{Zaslavsky_2023}
{Zaslavsky}, A. 2023, \grl, 50, e2022GL101548

\bibitem[\protect\citeauthoryear{Zhao et~al.}{Zhao et~al.}{2021}]{Zhao_2021}
Zhao, L.-L., Zank, G.~P., He, J.~S., Telloni, D., Adhikari, L., Nakanotani, M., Kasper, J.~C.,  \& Bale, S.~D. 2021, The Astrophysical Journal, 922, 188

\end{thebibliography}
\end{document}